\documentclass[
superscriptaddress,
showpacs,preprintnumbers,
 amsmath,amssymb,
 aps,
 twocolumn,
prl,
footinbib
]{revtex4-1}
\usepackage{braket}
\usepackage{bm}
\usepackage{dcolumn}
\usepackage{cancel}
\usepackage[dvipdfmx,final]{graphicx}
\input{colordvi.tex}
\usepackage{longtable}
\usepackage[compat=1.1.0]{tikz-feynman}
\usepackage{tabularx}
\usepackage{cleveref}
\tikzfeynmanset{every edge={very thick},}
\def\bea{\begin{eqnarray}}
\def\eea{\end{eqnarray}}

\newcommand\unit[1]{\,{\rm #1}}

\def\kpc{\unit{kpc}}
\def\pc{\unit{pc}}

\usepackage{color}

\begin{document}
\preprint{CTPU-PTC-19-33}
\title{Escalating core formation with dark matter self-heating}
\author{Ayuki Kamada}
\email{akamada@ibs.re.kr}
\affiliation{Center for Theoretical Physics of the Universe, Institute for Basic Science (IBS), Daejeon 34126, Korea}
\author{Hee Jung Kim}
\email{hyzer333@kaist.ac.kr}
\affiliation{Department of Physics, KAIST, Daejeon 34141, Korea}
\date{\today}

\begin{abstract}
Exothermic scatterings of dark matter (DM) produce DM particles with significant kick velocities inside DM halos.
In collaboration with DM self-interaction, the excess kinetic energy of the produced DM particles is distributed to the others, which self-heats the DM particles as a whole.
The DM self-heating is efficient towards the halo center, and the heat injection is used to enhance the formation of a uniform density core inside halos.
The effect of DM self-heating is expected to be more significant in smaller halos for two reasons:
1) the exothermic cross section times the relative velocity, $\left\langle\sigma_{\rm exo}v_{\rm rel}\right\rangle$, is constant;
2) and the ratio of the injected heat to the velocity dispersion squared gets larger towards smaller halos.
For the first time, we quantitatively investigate the core formation from DM self-heating for halos in a wide mass range ($10^{9}$--$10^{15}\,{\rm M}_\odot$) using the gravothermal fluid formalism.
Notably, we demonstrate that the core formation is sharply escalating towards smaller halos by taking the self-heating DM (i.e., DM that semi-annihilates and self-interacts) as an example.
We show that the sharp escalation of core formation may cause a tension in simultaneously explaining the observed central mass deficit of Milky Way satellites, and field dwarf/low surface brightness spiral galaxies.
While the details of the self-heating effect may differ among models, we expect that the sharp halo-mass dependence of the core formation is a general feature of exothermic DM.
\end{abstract}

\maketitle

\paragraph{\underline{Introduction}.---}
Dark matter (DM) shapes the matter distribution of our Universe.
It provides gravitational potential wells where galaxies are formed to evolve and merge.
While their existence and distribution is probed through their gravitation, the particle nature of DM remains unknown.
Perhaps one of the most motivated candidates is a weakly interacting massive particle (WIMP), and there have been extensive efforts to detect them by probing their interactions with the Standard Model (SM).
However, there has been no conclusive signal yet~\cite{Arcadi:2017kky,Roszkowski:2017nbc}.

In light of this situation, there have been a great deal of efforts to look for new benchmarks for DM.
Many of the attempts in this direction invoke new mechanisms that stabilize DM,
which accompany additional particle contents and DM interactions within a dark sector.
Such secluded DM interactions may leave an imprint on the distribution of DM in our Universe~\cite{Buckley:2017ijx}.
Thus, looking for distinctive consequences of DM interactions on their spatial distribution could provide hints
that may be difficult to achieve through traditional particle physics probes, i.e., large hadron collider (LHC)~\cite{Abercrombie:2015wmb,Aaboud:2017phn,Sirunyan:2017jix} and direct/indirect detection searches~\cite{Aprile:2018dbl,Cui:2017nnn,Ahnen:2016qkx}.

The discrepancies between the predictions of the standard CDM paradigm and the astrophysical observations in small-scale structures~\cite{Bullock:2017xww} (small-scale issues) may be hinting the existence of dark sector interactions.
While it is not clear if sub-grid astrophysical processes could entirely resolve the issues~\cite{DiCintio:2013qxa,DiCintio:2014xia,Oman:2015xda},
it is an interesting possibility that the secluded DM interactions are responsible for the discrepancies.
A prominent aspect of the issues is that observed galaxies prefer cored central density profile, rather than cuspy~\cite{Spergel:1999mh,Firmani:2000ce}.
Such central mass deficit is reported to be halo-mass dependent;
galaxy clusters ($10^{14}$--$10^{15}\,{\rm M}_\odot$)~\cite{Sand:2003bp,Newman:2012nw} exhibit milder cores than 
dwarf spheroidal galaxies in the field~\cite{Flores:1994gz,Walker:2011zu} and low surface brightness (LSB) spiral galaxies~\cite{deBlok:2001hbg,deBlok:2002vgq,Simon:2004sr} ($10^{10}$--$10^{11}\,{\rm M}_\odot$).

Self-interacting DM (SIDM) is one of the well known framework for explaining the central mass deficit of halos~\cite{Tulin:2017ara},
and the halo-mass dependent cores may be explained by the velocity dependent self-interaction of DM particles~\cite{Kaplinghat:2015aga}.
One of the realizations for the velocity dependence is the exothermic self-interaction; kinematically unsuppressed exothermic cross section exhibits $\sigma_{\rm exo}\propto1/v_{\rm rel}$ in the non-relativistic limit~\cite{McDermott:2017vyk}.
Meanwhile, the energy released from such exothermic DM scatterings give rise to yet another scale dependent effect:
the impact of the released energy increases towards smaller halos~\cite{Loeb:2010gj,Schutz:2014nka}.
The injected energy self-heats the DM particles, and has been shown to enhance the core formation towards smaller halos through analytic estimations~\cite{Chu:2018nki} and $N$-body simulations~\cite{Vogelsberger:2018bok}.
On the other hand, there has been no concrete study that scopes the effects of DM self-heating in a wide mass range of halos.

\begin{figure*}
\centering
\includegraphics[scale=0.6]{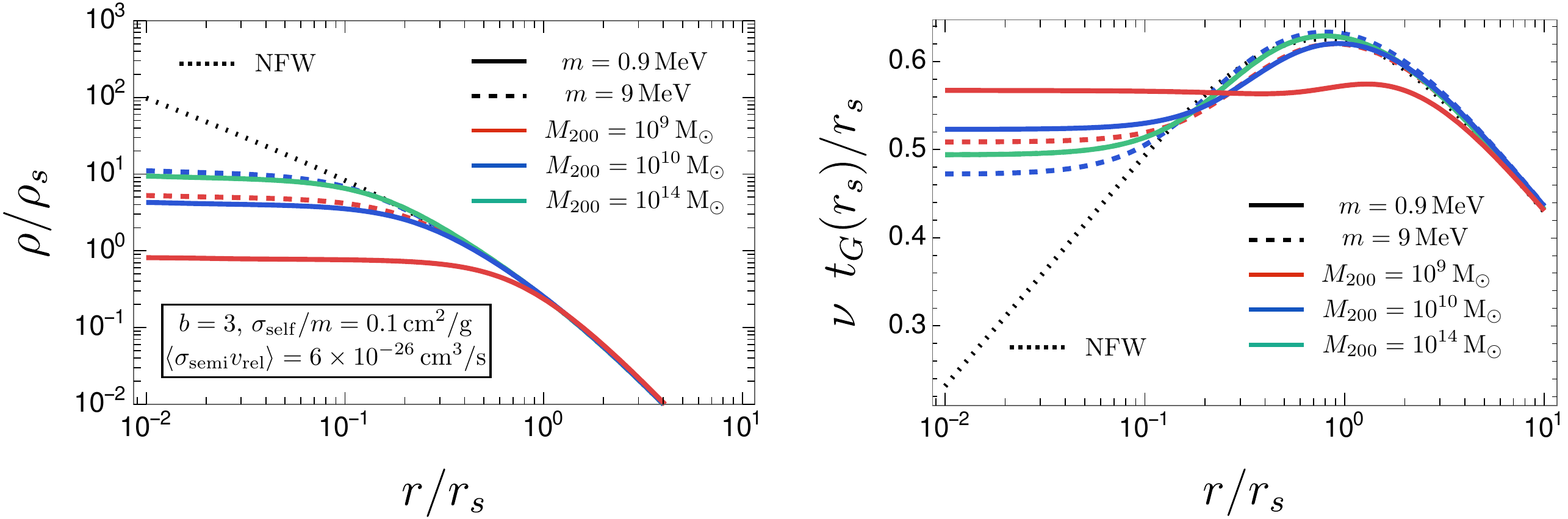}
\caption{
{\it Left panel} - Resultant density profiles of SHDM halos.
We took the evolution time that is determined by the virial mass of a halo~\cite{Note1}.
Colored lines represent different masses of halos;
a galaxy cluster (green), a field dwarf/LSB galaxy (blue), and a MW satellite galaxy (red).
NFW profile (dotted black) is used as initial condition;
we use the concentration-mass relation from Ref.~\cite{Dutton:2014xda} to determine $\rho_s$ and $r_s$.
We use the fudge factor $b=3$ (see Eq.~\eqref{eq:xi}) to match the results of the $N$-body simulation for inelastic SIDM~\cite{Note1}.
Solid lines represent the SHDM mass ($m=0.9\,{\rm MeV}$) that reproduce similar core size as the inferred pure SIDM cross section for field dwarf/LSB galaxies, $\sigma_{\rm self}/m=1.9^{+0.6}_{-0.4}\,{\rm cm^2/g}$~\cite{Kaplinghat:2015aga}.
The core formation is efficient towards smaller halos.
For galaxy clusters, the resultant profiles do not depend much on the SHDM mass;
the effect of DM self-heating is negligible, and the core formation is predominantly determined by the SIDM cross section.
{\it Right panel} - Same as the left panel but for the 1-dimensional velocity dispersion. $t_{G}(r_s)=1/\sqrt{4\pi G\rho_{\rm NFW}(r_s)}$ is the gravitational time scale with $\rho_{\rm NFW}(r_s)=\rho_s/4$.
}
\label{fig:sample}
\end{figure*}

In this work, we investigate the impact of DM self-heating on isolated halos by adopting the gravothermal fluid formalism~\cite{1980MNRAS.191..483L,GalacticDynamics}.
The fluid formalism allows us to follow the evolutions of halos as small as the Milky Way (MW) satellites ($10^9\,{\rm M}_\odot$) at a small computational expense, which is difficult to achieve in $N$-body simulations.
Exploiting the method, we investigate the effect of DM self-heating on halos in a wide mass range ($10^{9}$--$10^{15}\,{\rm M}_\odot$) for the first time.
To demonstrate the halo-mass dependence of core formation from the self-heating effect, we take self-heating DM (SHDM)~\cite{Kamada:2017gfc,Kamada:2018hte} as an example, where DM particles semi-annihilate (${\rm DM}+{\rm DM}\rightarrow{\rm DM}+\phi$, where $\phi$ is a light particle) and self-interact with {\it constant} $\left\langle\sigma_{\rm semi} v_{\rm rel}\right\rangle$ and $\sigma_{\rm self}/m$.
Interestingly, we find that the core formation effect from DM self-heating is sharply escalated for halos smaller than a certain mass.
We demonstrate that this sharp halo-mass dependence may cause a tension between parameters that explain the central mass deficit of MW satellite dwarf spheroidal galaxies~\cite{Gilmore:2007fy} and those of field dwarf/LSB galaxies, although the mass difference between the two is just an order of magnitude.
We discuss that the sharp escalation of core formation may be present in other models of exothermic DM~\cite{McDermott:2017vyk,Schutz:2014nka,Vogelsberger:2018bok}, and attentive investigations on this feature would be needed.
\\
\paragraph{\underline{Modeling dark matter self-heating}.---}

We study the effect of DM self-heating by following the gravothermal evolution of isolated halos that consist of SHDM.
We numerically solve the gravothermal fluid equations that are analogous to the ones adopted to describe halos of pure SIDM~\cite{Balberg:2001qg,Ahn:2004xt,Koda:2011yb,Pollack:2014rja,Essig:2018pzq,Nishikawa:2019lsc},
but with an additional heat injection term that models the DM self-heating.
We first review the gravothermal fluid model for the pure SIDM, and discuss our implementation of DM self-heating.

We consider a spherically symmetric initial halo that is formed around the time $t_\ast$ that is determined by its virial mass, based on a spherical collapse model~\cite{Note1}.
We {\it assume} that DM self-interaction and DM self-heating are unimportant during the process of halo formation;
we use the initial halo profiles predicted from CDM $N$-body simulations~\cite{Note1}.
Then the initial halo density profile is approximated by the Navarro-Frenk-White (NFW) profile~\cite{Navarro:1995iw,Navarro:1996gj},
$\rho_{\rm NFW}(r)=\rho_s /[(r/r_s)(1+(r/r_s))^2]$,
where $r_s$ ($\rho_s$) is the NFW scale radius (density).
We use the NFW scale parameters determined from the halo virial mass ($M_{200}$), using the concentration-mass relation inferred from CDM $N$-body simulations~\cite{Dutton:2014xda}: $\rho_s\simeq 0.011\,{\rm M}_{\odot}/{\rm pc}^3\,(10^{10}\,{\rm M}_{\odot}/M_{200})^{0.24}$, $r_s\simeq 3.43\,{\rm kpc}\, (M_{200}/10^{10}\,{\rm M}_{\odot})^{0.44}$.

The gravothermal fluid formalism approximates the DM particles as ideal fluid that is described by their mass density $\rho$ and fluid pressure $p$.
The initial halo is assumed isotropic in pressure, and we define the 1-dimensional velocity dispersion as $\nu(r,t)=\sqrt{p(r,t)/\rho(r,t)}$.
In the case of pure SIDM, gravothermal evolution of DM fluid is governed by the following equations~\cite{Balberg:2001qg,Ahn:2004xt,Koda:2011yb,Pollack:2014rja,Essig:2018pzq,Nishikawa:2019lsc}:
\begin{subequations}
\label{eq:gravothermaleqns}
    \noindent\begin{minipage}{0.27\textwidth}
    \begin{equation}
\frac{\partial\left(\rho\nu^{2}\right)}{\partial r}+\frac{GM\rho}{r^{2}}=0\label{eq:1Euler}\,,
\end{equation}
    \end{minipage}%
    \begin{minipage}{0.21\textwidth}
    \begin{equation}
\frac{\partial M}{\partial r}=4\pi r^{2}\rho\,,\label{eq:1mass}
\end{equation}
    \end{minipage}%
    \\
     \begin{minipage}{0.45\textwidth}
\begin{equation}
\frac{3}{\nu}\left(\frac{\partial\nu}{\partial t}\right)_{M}-\frac{1}{\rho}\left(\frac{\partial\rho}{\partial t}\right)_{M}=\frac{1}{\nu^{2}}\frac{\delta u_{\rm cond}}{\delta t}\label{eq:1entropy}\,,
\end{equation}
    \end{minipage}\vskip1em
\end{subequations}
\noindent where $M(r,t)$ is the fluid mass enclosed within radius $r$, and $G$ is the Newton's constant.
$(\partial_t)_M=\partial_t+\boldsymbol{V}\cdot\nabla$ is the Lagrangian time derivative where $\boldsymbol{V}$ is the fluid bulk velocity; it refers to changes within the fluid element as it changes its state and location.
We note that $\boldsymbol{V}$ is determined by the continuity equation, $\partial_t\rho+\nabla\cdot(\rho\boldsymbol{V})=0$.
Meanwhile, the terms associating $\boldsymbol{V}$ in the fluid momentum conservation equation (Euler equation) is assumed to be negligible [Eq.~\eqref{eq:1Euler}];
we require the halo to be in quasi-static hydrostatic equilibrium~\cite{Note1}.
Eq.~\eqref{eq:1mass} states that the mass of DM fluid is virtually conserved.

Eq.~\eqref{eq:1entropy} is the first law of thermodynamics.
The details of particle physics appear in the RHS, which is the rate of heat gain of a fluid element;
fluid elements conduct heat with neighboring sites in the radial direction through DM self-interaction.
The heat conduction is modeled with a heat diffusion equation:
\begin{equation}
\frac{\delta u_{\rm cond}}{\delta t}=\frac{m}{\rho}\nabla\cdot\left(\kappa \nabla \nu^2\right)\,,
\end{equation}
where $u$ is the specific energy (energy per DM mass), and $\kappa$ is the thermal conductivity.
We follow the implementation of $\kappa$ in Refs.~\cite{Note1,Balberg:2001qg,Koda:2011yb}.
The net heat gain from heat conduction is used to increase the velocity dispersion (the first term of the LHS), and to decrease the mass density (the second term of the LHS) by expanding the volume of the DM fluid.

We implement the effects of DM self-heating by introducing another heating term in the RHS of Eq.~\eqref{eq:1entropy}.
In the case of SHDM, 
a tiny fraction of the boosted DM particles produced from DM semi-annihilations is captured before escaping the halo through DM self-interaction.
(For simplicity, we assume that the produced $\phi$'s escape from the halo and plays no role in halo evolution.)
The high kinetic energy of the captured boosted DM particle, $\delta E=m/4$, is then distributed to the others, i.e., DM particles self-heat.
We effectively describe this process as injecting heat to the local DM fluid element with the following rate:
\begin{equation}
\frac{\delta u_{\rm semi}}{\delta t}=\frac{\rho \left\langle\sigma_{\rm semi} v_{\rm rel}\right\rangle}{m}\frac{\xi\delta E}{m}\,,
\label{eq:semiinjection}
\end{equation}
where $\xi$ is the efficiency to capture a boosted DM particle.
The realistic modeling of the capture efficiency $\xi$ is non-trivial;
the production site of the captured boosted DM particle may not be the local fluid element, but some remote site from the captured region.
Instead, we use a constant and uniform profile for $\xi$:
\begin{equation}
\begin{aligned}
\xi =& \,b\times \frac{r}{\lambda}\big|_{r=r_s} \\
\simeq&\,0.0002\,b\,\left(\frac{r_s}{3.43\,\kpc}\right)\\
&\times\left(\frac{\rho_s}{0.011\,M_{\odot}/{\rm pc}^{3}}\right)\left(\frac{\sigma_{{\rm self}}/m}{0.1\,{\rm cm^{2}}/{\rm g}}\right)\,,
\label{eq:xi}
\end{aligned}
\end{equation}
where $b$ is a fudge factor and $\lambda= 1/(\rho \sigma_{\rm self}/m)$ is the free-streaming length of a DM particle.
Hereafter, we set the fudge factor to be $b=3$;
this choice of the fudge factor is motivated from the $N$-body simulation result for the inelastic SIDM~\cite{Note1}. 
For other exothermic DM models, we may replace the parameters $\left\langle\sigma_{\rm exo}v_{\rm rel}\right\rangle/m\times\delta E/m$ in Eq,~\eqref{eq:semiinjection} with the corresponding values.
Taking into account the DM self-heating represented by Eq.~\eqref{eq:semiinjection}, we numerically solve Eqs.~\eqref{eq:gravothermaleqns} using the methods described in Ref.~\cite{Balberg:2002ue,Pollack:2014rja,Essig:2018pzq,Nishikawa:2019lsc}.
We find that DM self-heating leads to a formation of uniform cores inside halos (see FIG.~\ref{fig:sample}).
\\
\paragraph{\underline{Halo mass dependence of core formation}.---}
The core formation from self-heating effect becomes stronger towards smaller halos.
This is understood by observing the DM self-heating term in the RHS of Eq.~\eqref{eq:1entropy}:
\begin{equation}
\frac{1}{\nu^2}\frac{\delta u_{\rm semi}}{\delta t}
=\left(\frac{\nu_{\rm NFW}(r_s)}{\nu}\right)^2\left(\frac{\rho}{\rho_{\rm NFW}(r_s)}\right)\frac{1}{t_{\rm heat}}\,,
\label{eq:semiheat}
\end{equation}
where $\nu_{\rm NFW}(r_s)\simeq0.3\sqrt{4\pi G \rho_s r_s^2}$ and $\rho_{\rm NFW}(r_s)=\rho_s/4$ are respectively the velocity dispersion and density of the initial NFW halo at the scale radius $r=r_s$.
$t_{\rm heat}$ is the representative heating time scale of the initial NFW halo;
it is roughly the time scale required for a DM particle to absorb energy comparable to its kinetic energy $\sim m\nu^2$ by capturing boosted DM particles produced from semi-annihilation:
\begin{equation}
\begin{aligned}
t_{\rm heat} =&\,\left(\frac{\rho_{\rm NFW}(r_s)\left\langle \sigma_{{\rm semi}}v_{{\rm rel}}\right\rangle }{m}\frac{\xi\delta E}{m\nu_{\rm NFW}^2(r_s)}\right)^{-1}\\
\simeq&\,46\,{\rm Gyr}\,\left(\frac{M_{200}}{10^{9}\,{\rm M}_{\odot}}\right)^{0.68}\left(\frac{0.1\,{\rm cm^{2}/g}}{\sigma_{{\rm self}}/m}\right)\\
&\times\left(\frac{6\times10^{-26}{\rm cm^{3}/s/{\rm MeV}}}{\left\langle \sigma_{{\rm semi}}v_{{\rm rel}}\right\rangle/m}\right)\left(\frac{1/4}{\delta E/m}\right)\,,
\end{aligned}
\label{eq:tJ}
\end{equation}
where we took $b=3$ and $M_{200}$ is the virial mass of a halo.
In the second equality, we used Eq.~\eqref{eq:xi}, and the concentration-mass relation~\cite{Dutton:2014xda} for the NFW scale parameters.
In Eq.~\eqref{eq:semiheat}, $\nu(r)/\nu_{\rm NFW}(r_s)$ and $\rho(r)/\rho_{\rm NFW}(r_s)$ for given $r/r_s$ do not depend much on halo mass, while $t_{\rm heat}$ is significantly shorter in smaller halos.
Thus, the core formation from self-heating is increasing towards smaller halos.
This feature is illustrated in FIG.~\ref{fig:sample}.

\begin{figure*}
\centering
\includegraphics[scale=0.6]{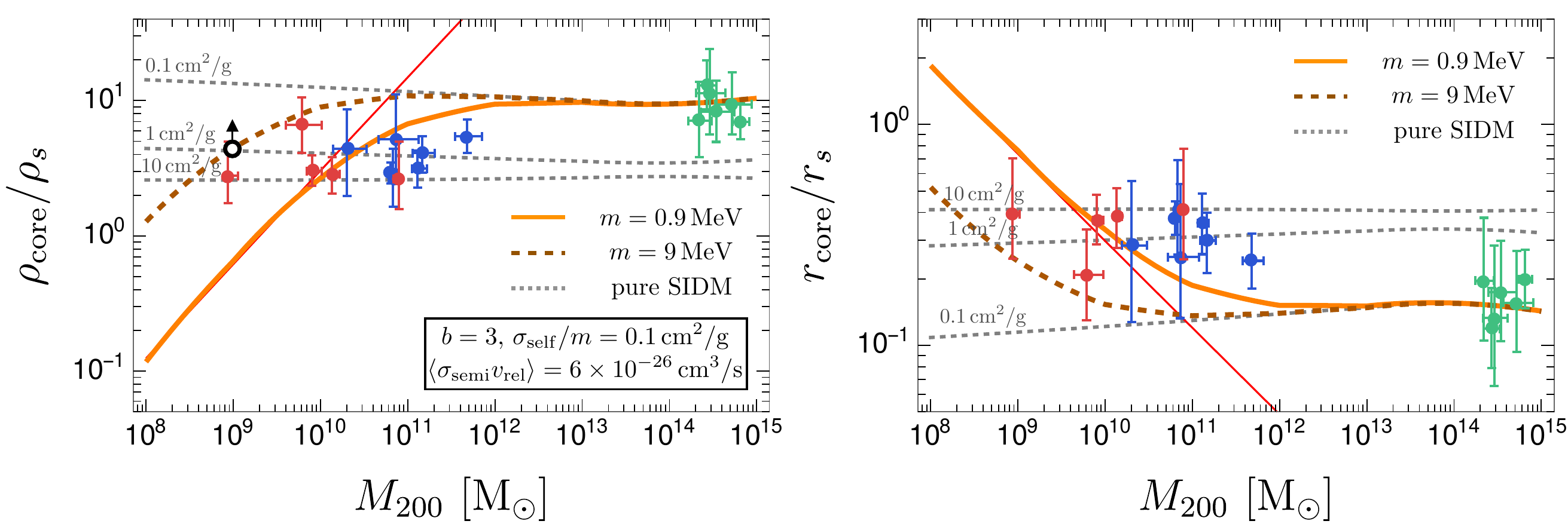}
\caption{
{\it Left panel} - Halo-mass dependence of resultant core densities of SHDM halos (in orange).
We took the evolution time that is determined by the virial mass of a halo~\cite{Note1}.
Gray curves represent the case of pure SIDM.
The data points are the core densities predicted from the inferred values of pure $\sigma_{\rm self}/m$ from the observations on field dwarf(red)/LSB(blue) galaxies, and galaxy clusters (green)~\cite{Kaplinghat:2015aga}.
The core formation is sharply escalating towards smaller halos, which appears to disfavor the SHDM as a solution to small-scale issues;
while the SHDM mass $m=0.9\,{\rm MeV}$ (solid orange) is preferred to explain the core size of field dwarf/LSB galaxies, the central densities of MW satellites~\cite{Gilmore:2007fy} (black circle) require the SHDM mass to be heavier than $9\,{\rm MeV}$ (dashed orange).
This is in contrast to the pure SIDM;
there is a lower limit on the core density since the SIDM halos undergo the gravothermal collapse at present for $\sigma_{\rm self}/m\gtrsim 10\,{\rm cm^2/g}.$
The red line represents the analytic estimation given in Eq.~\eqref{eq:denestimate}.
{\it Right panel} - Same as the left panel but for the resultant core radius. The red line represents the analytic estimation for $r_{\rm core}$ in the self-heating dominated limit; $r_{\rm core}(t_{\rm age})\simeq 1.4\times r_s\left((t_{\rm age}-t_\ast)/t_{\rm heat}\right)^{0.57}$.
}
\label{fig:eff}
\end{figure*}

This halo-mass dependence of core formation may be used to explain why the observed central mass deficit is more appreciable in smaller halos.
In FIG.~\ref{fig:eff}, we demonstrate that core formation of (sub-)${\rm MeV}$ scale SHDM may address this issue even with constant $\left\langle\sigma_{\rm semi}v_{\rm rel}\right\rangle\simeq6\times10^{-26}\,{\rm cm^3/s}$ and $\sigma_{\rm self}/m\simeq 0.1\,{\rm cm^2/g}$.
There, we define the core density $\rho_{\rm core}$ to be the density at the innermost radius, and the core radius $r_{\rm core}$ to be the radius where the density is $\rho(r_{\rm core})=\rho_{\rm core}/2$.
We find that the SHDM mass $m\simeq 0.9\,{\rm MeV}$ (solid orange curve in FIG.~\ref{fig:eff}) is preferred to explain the observed core density/size of dwarf/LSB galaxies.
We take these parameters as a benchmark for SHDM.
\\
\paragraph{\underline{Constraints from Milky Way satellites}.---}
As illustrated in FIG.~\ref{fig:sample} and FIG.~\ref{fig:eff}, the core formation from self-heating effect is stronger towards smaller halos, which may address why galaxy clusters exhibit milder cores than dwarf/LSB galaxies.
On the other hand, the core formation from self-heating in smaller halos ($M_{200}\lesssim 10^{10}\,{\rm M}_\odot$) may be too efficient to be compatible with observations.
In fact, observed central densities of MW satellite dwarf spheroidal galaxies provide strong restrictions on the efficiency of the self-heating.

For the benchmark parameters of SHDM, the core formation of MW satellites is basically dominated by the self-heating effect from semi-annihilation: $t_{\rm heat}<t_{\rm self}(r=r_s)$, where $t_{\rm self}=1/(a\rho\nu\sigma_{\rm self}/m)$ is the self-interaction time scale with $a=\sqrt{16/\pi}$.
In such a case, the following is a good estimation for the core density~\cite{Note2}:
\begin{equation}
\rho_{\rm core}(t_{\rm age})\simeq0.2\times\rho_s \left(\frac{t_{\rm heat}}{t_{\rm age}-t_\ast}\right)\,,
\label{eq:denestimate}
\end{equation}
where $t_\ast$ is the time when the initial virialized halo was formed, and $t_{\rm age}\simeq13.8\,{\rm Gyr}$ is the age of the Universe.
The observed central densities of MW satellite dwarf galaxies is in the range of $\rho_{\rm c, obs}\sim0.1$--$1\,{\rm M}_{\odot}/{\rm pc}^3$~\cite{Gilmore:2007fy}.
We require the predicted core density to be larger than the observed one ($\rho_{\rm core}(t_{\rm age})\gtrsim0.1\,{\rm M}_{\odot}/{\rm pc}^3$), 
which restricts the heating time scale for MW satellites:
\begin{equation}
t_{\rm heat}\gtrsim320\,{\rm Gyr}\,\left(\frac{\rho_{\rm c,obs}}{0.1\,{\rm M}_{\odot}/{\rm pc}^3}\right)\left(\frac{M_{200,{\rm infall}}}{10^9\,{\rm M_\odot}}\right)^{0.24}\,.
\label{eq:MWconstraint}
\end{equation}
Here we took the masses of MW satellites as their masses prior to accretion onto the MW, which is estimated using the concentration-mass relation of field halos~\cite{Wolf:2009tu}.
Remark that our choice of a satellite halo mass is conservative; using their mass at present would provide stronger restriction to the heating efficiency, since progenitor halos of MW satellite galaxies lose their mass along the accretion.
With the benchmark cross sections, this provides a rough lower bound on SHDM mass: $m\gtrsim 9\,{\rm MeV}$.
The core formation for SHDM with mass of $m=9\,{\rm MeV}$ is shown in FIG.~\ref{fig:eff} (dashed orange): for these parameters, the core formation may not be efficient enough to explain the inner density profiles of field dwarf/LSB galaxies.

It is interesting that observations on MW satellites provide a significant constraint, although their masses (at infall) are smaller than those of field dwarf/LSB galaxies by just an order of magnitude.
This is because the core formation from DM self-heating is sharply escalating towards smaller halos, which is represented by the sharp power-law halo-mass dependence of $t_{\rm heat}$ in Eq.~\eqref{eq:tJ}.
Thus, observations on even smaller halos (i.e., minihalos) would be a good probe to examine the self-heating nature of DM.
For example, the observed concentration-mass relation of minihalos may provide a firm test to distinguish SHDM from pure SIDM;
SHDM halos would exhibit consistently lower concentration than the pure SIDM halos.
This is because the core density of self-heating dominated halos decreases indefinitely until present, as in Eq.~\eqref{eq:denestimate}.
This is in contrast to pure SIDM halos, since their core density do not get lower than $\rho_{\rm NFW}(r_s)=\rho_s/4$ due to the self-similar gravothermal collapse~\cite{Balberg:2002ue}.
\\
\paragraph{\underline{Self-heating in other models}.---}
In the preceding sections, we discussed the core formation of a specific scenario, SHDM.
We demonstrated that the core formation of SHDM is sharply escalating towards smaller halos, which appears to disfavor SHDM as a solution to small-scale issues.
We expect that this escalating core formation is general among scenarios that exhibit DM self-heating.

Here we introduce other selected scenarios that exhibit DM self-heating;
dark nuclei as DM particles that undergo two-to-two fusion processes that convert DM mass deficit into kinetic energy (fusion DM)~\cite{McDermott:2017vyk};
two-state DM particles which are collisionally (de-)excited between nearly degenerate energy states (eXciting DM)~\cite{Schutz:2014nka,Vogelsberger:2018bok}.
Since both models exhibit comparable DM elastic self-scattering, we may use the modeling for the heat injection as in Eq.~\eqref{eq:semiinjection};
we replace $\left\langle\sigma_{\rm semi}v_{\rm rel}\right\rangle/m\times \delta E/m$ with the corresponding values in other models.
The heating time scales for MW satellites with the suggested benchmark parameters are given as
\begin{equation}
t_{{\rm heat}}\simeq\begin{cases}
5\,{\rm Gyr}\,\left(\frac{147\,{\rm cm^{2}/g\times km/s}}{\left\langle \sigma_{{\rm fusion}}v_{{\rm rel}}\right\rangle /m}\right)\left(\frac{0.1\,{\rm cm^{2}/g}}{\sigma_{{\rm self}}/m}\right)\left(\frac{10^{-5}}{\delta E/m}\right)\\
\textrm{\qquad \qquad \qquad \qquad \qquad \quad for ``fusion" DM,}\\
5\,{\rm Gyr}\,\left(\frac{20\,{\rm cm^{2}/g\times km/s}}{\left\langle \sigma_{{\rm ex\rightarrow gr}}v_{{\rm rel}}\right\rangle /m}\right)\left(\frac{2\,{\rm cm^{2}/g}}{\sigma_{{\rm self}}/m}\right)\left(\frac{2\times10^{-6}}{\delta E/m}\right)\\
\textrm{\qquad \qquad \qquad \qquad \qquad  for ``eXciting" DM,}
\end{cases}
\label{eq:exomodels}
\end{equation}
where we took $b=3$, and assumed degenerate final state DM masses for the fusion DM.
Although both models exhibit large $\left\langle\sigma_{\rm exo}v_{\rm rel}\right\rangle/m$, the energy release from an exothermic process $\delta E/m$ is small, so that the heating timescales are similar to that of SHDM [Eq.~\eqref{eq:tJ}].
The benchmark heating time scales of the two models are shorter than the limit of Eq.~\eqref{eq:MWconstraint}.
Therefore, the possible parameters space that is consistent with central densities of MW satellites needs to be carefully examined.
\\
\paragraph{\underline{Conclusions}.---}
Using the gravothermal fluid formalism, we have studied the impact of DM self-heating on halo structures.
We took SHDM as an example, and showed that DM self-heating flattens the central cuspy profile of an initial halo and develops a uniform core.
The fluid formalism allowed us to scope the effect of DM self-heating for a wide mass range of halos,
and we found that the core formation is sharply escalating towards smaller halos.
By requiring the SHDM to be consistent with the observed central densities of MW satellites, we showed that the escalating core formation of SHDM appears to be disfavored as an explanation for the observed central mass deficit of field dwarf/LSB galaxies.
We estimated the heating time scales in other exothermic DM scenarios which are invoked to resolve the small-scale issues,
and demonstrated that careful investigations are needed to identify the parameter region consistent with central densities of MW satellites.
\\
\paragraph{\underline{Acknowledgments}.---} 
We would like to thank Hitoshi Murayama for drawing our attention to this project.
We also thank Hyungjin Kim for the detailed comments on the manuscript.
A. K. would like to acknowledge the Mainz Institute for Theoretical Physics (MITP) of the Cluster of Ex-cellence PRISMA+ (Project ID 39083149) for enabling A. K. to complete a significant portion of this work. The work of A. K. is supported by IBS under the project code, IBS-R018-D1.

\bibliography{SelfheatDM}

\begin{thebibliography}{52}
\expandafter\ifx\csname natexlab\endcsname\relax\def\natexlab#1{#1}\fi
\expandafter\ifx\csname bibnamefont\endcsname\relax
  \def\bibnamefont#1{#1}\fi
\expandafter\ifx\csname bibfnamefont\endcsname\relax
  \def\bibfnamefont#1{#1}\fi
\expandafter\ifx\csname citenamefont\endcsname\relax
  \def\citenamefont#1{#1}\fi
\expandafter\ifx\csname url\endcsname\relax
  \def\url#1{\texttt{#1}}\fi
\expandafter\ifx\csname urlprefix\endcsname\relax\def\urlprefix{URL }\fi
\providecommand{\bibinfo}[2]{#2}
\providecommand{\eprint}[2][]{\url{#2}}

\bibitem[{\citenamefont{Arcadi et~al.}(2018)\citenamefont{Arcadi, Dutra, Ghosh,
  Lindner, Mambrini, Pierre, Profumo, and Queiroz}}]{Arcadi:2017kky}
\bibinfo{author}{\bibfnamefont{G.}~\bibnamefont{Arcadi}},
  \bibinfo{author}{\bibfnamefont{M.}~\bibnamefont{Dutra}},
  \bibinfo{author}{\bibfnamefont{P.}~\bibnamefont{Ghosh}},
  \bibinfo{author}{\bibfnamefont{M.}~\bibnamefont{Lindner}},
  \bibinfo{author}{\bibfnamefont{Y.}~\bibnamefont{Mambrini}},
  \bibinfo{author}{\bibfnamefont{M.}~\bibnamefont{Pierre}},
  \bibinfo{author}{\bibfnamefont{S.}~\bibnamefont{Profumo}}, \bibnamefont{and}
  \bibinfo{author}{\bibfnamefont{F.~S.} \bibnamefont{Queiroz}},
  \bibinfo{journal}{Eur. Phys. J.} \textbf{\bibinfo{volume}{C78}},
  \bibinfo{pages}{203} (\bibinfo{year}{2018}), \eprint{1703.07364}.

\bibitem[{\citenamefont{Roszkowski et~al.}(2018)\citenamefont{Roszkowski,
  Sessolo, and Trojanowski}}]{Roszkowski:2017nbc}
\bibinfo{author}{\bibfnamefont{L.}~\bibnamefont{Roszkowski}},
  \bibinfo{author}{\bibfnamefont{E.~M.} \bibnamefont{Sessolo}},
  \bibnamefont{and}
  \bibinfo{author}{\bibfnamefont{S.}~\bibnamefont{Trojanowski}},
  \bibinfo{journal}{Rept. Prog. Phys.} \textbf{\bibinfo{volume}{81}},
  \bibinfo{pages}{066201} (\bibinfo{year}{2018}), \eprint{1707.06277}.

\bibitem[{\citenamefont{Buckley and Peter}(2018)}]{Buckley:2017ijx}
\bibinfo{author}{\bibfnamefont{M.~R.} \bibnamefont{Buckley}} \bibnamefont{and}
  \bibinfo{author}{\bibfnamefont{A.~H.~G.} \bibnamefont{Peter}},
  \bibinfo{journal}{Phys. Rept.} \textbf{\bibinfo{volume}{761}},
  \bibinfo{pages}{1} (\bibinfo{year}{2018}), \eprint{1712.06615}.

\bibitem[{\citenamefont{Abercrombie et~al.}(2015)}]{Abercrombie:2015wmb}
\bibinfo{author}{\bibfnamefont{D.}~\bibnamefont{Abercrombie}}
  \bibnamefont{et~al.} (\bibinfo{year}{2015}), \eprint{1507.00966}.

\bibitem[{\citenamefont{Aaboud et~al.}(2018)}]{Aaboud:2017phn}
\bibinfo{author}{\bibfnamefont{M.}~\bibnamefont{Aaboud}} \bibnamefont{et~al.}
  (\bibinfo{collaboration}{ATLAS}), \bibinfo{journal}{JHEP}
  \textbf{\bibinfo{volume}{01}}, \bibinfo{pages}{126} (\bibinfo{year}{2018}),
  \eprint{1711.03301}.

\bibitem[{\citenamefont{Sirunyan et~al.}(2018)}]{Sirunyan:2017jix}
\bibinfo{author}{\bibfnamefont{A.~M.} \bibnamefont{Sirunyan}}
  \bibnamefont{et~al.} (\bibinfo{collaboration}{CMS}), \bibinfo{journal}{Phys.
  Rev.} \textbf{\bibinfo{volume}{D97}}, \bibinfo{pages}{092005}
  (\bibinfo{year}{2018}), \eprint{1712.02345}.

\bibitem[{\citenamefont{Aprile et~al.}(2018)}]{Aprile:2018dbl}
\bibinfo{author}{\bibfnamefont{E.}~\bibnamefont{Aprile}} \bibnamefont{et~al.}
  (\bibinfo{collaboration}{XENON}), \bibinfo{journal}{Phys. Rev. Lett.}
  \textbf{\bibinfo{volume}{121}}, \bibinfo{pages}{111302}
  (\bibinfo{year}{2018}), \eprint{1805.12562}.

\bibitem[{\citenamefont{Cui et~al.}(2017)}]{Cui:2017nnn}
\bibinfo{author}{\bibfnamefont{X.}~\bibnamefont{Cui}} \bibnamefont{et~al.}
  (\bibinfo{collaboration}{PandaX-II}), \bibinfo{journal}{Phys. Rev. Lett.}
  \textbf{\bibinfo{volume}{119}}, \bibinfo{pages}{181302}
  (\bibinfo{year}{2017}), \eprint{1708.06917}.

\bibitem[{\citenamefont{Ahnen et~al.}(2016)}]{Ahnen:2016qkx}
\bibinfo{author}{\bibfnamefont{M.~L.} \bibnamefont{Ahnen}} \bibnamefont{et~al.}
  (\bibinfo{collaboration}{MAGIC, Fermi-LAT}), \bibinfo{journal}{JCAP}
  \textbf{\bibinfo{volume}{1602}}, \bibinfo{pages}{039} (\bibinfo{year}{2016}),
  \eprint{1601.06590}.

\bibitem[{\citenamefont{Bullock and Boylan-Kolchin}(2017)}]{Bullock:2017xww}
\bibinfo{author}{\bibfnamefont{J.~S.} \bibnamefont{Bullock}} \bibnamefont{and}
  \bibinfo{author}{\bibfnamefont{M.}~\bibnamefont{Boylan-Kolchin}},
  \bibinfo{journal}{Ann. Rev. Astron. Astrophys.}
  \textbf{\bibinfo{volume}{55}}, \bibinfo{pages}{343} (\bibinfo{year}{2017}),
  \eprint{1707.04256}.

\bibitem[{\citenamefont{Di~Cintio
  et~al.}(2014{\natexlab{a}})\citenamefont{Di~Cintio, Brook, Macciò, Stinson,
  Knebe, Dutton, and Wadsley}}]{DiCintio:2013qxa}
\bibinfo{author}{\bibfnamefont{A.}~\bibnamefont{Di~Cintio}},
  \bibinfo{author}{\bibfnamefont{C.~B.} \bibnamefont{Brook}},
  \bibinfo{author}{\bibfnamefont{A.~V.} \bibnamefont{Macciò}},
  \bibinfo{author}{\bibfnamefont{G.~S.} \bibnamefont{Stinson}},
  \bibinfo{author}{\bibfnamefont{A.}~\bibnamefont{Knebe}},
  \bibinfo{author}{\bibfnamefont{A.~A.} \bibnamefont{Dutton}},
  \bibnamefont{and} \bibinfo{author}{\bibfnamefont{J.}~\bibnamefont{Wadsley}},
  \bibinfo{journal}{Mon. Not. Roy. Astron. Soc.}
  \textbf{\bibinfo{volume}{437}}, \bibinfo{pages}{415}
  (\bibinfo{year}{2014}{\natexlab{a}}), \eprint{1306.0898}.

\bibitem[{\citenamefont{Di~Cintio
  et~al.}(2014{\natexlab{b}})\citenamefont{Di~Cintio, Brook, Dutton, Macciò,
  Stinson, and Knebe}}]{DiCintio:2014xia}
\bibinfo{author}{\bibfnamefont{A.}~\bibnamefont{Di~Cintio}},
  \bibinfo{author}{\bibfnamefont{C.~B.} \bibnamefont{Brook}},
  \bibinfo{author}{\bibfnamefont{A.~A.} \bibnamefont{Dutton}},
  \bibinfo{author}{\bibfnamefont{A.~V.} \bibnamefont{Macciò}},
  \bibinfo{author}{\bibfnamefont{G.~S.} \bibnamefont{Stinson}},
  \bibnamefont{and} \bibinfo{author}{\bibfnamefont{A.}~\bibnamefont{Knebe}},
  \bibinfo{journal}{Mon. Not. Roy. Astron. Soc.}
  \textbf{\bibinfo{volume}{441}}, \bibinfo{pages}{2986}
  (\bibinfo{year}{2014}{\natexlab{b}}), \eprint{1404.5959}.

\bibitem[{\citenamefont{Oman et~al.}(2015)}]{Oman:2015xda}
\bibinfo{author}{\bibfnamefont{K.~A.} \bibnamefont{Oman}} \bibnamefont{et~al.},
  \bibinfo{journal}{Mon. Not. Roy. Astron. Soc.}
  \textbf{\bibinfo{volume}{452}}, \bibinfo{pages}{3650} (\bibinfo{year}{2015}),
  \eprint{1504.01437}.

\bibitem[{\citenamefont{Spergel and Steinhardt}(2000)}]{Spergel:1999mh}
\bibinfo{author}{\bibfnamefont{D.~N.} \bibnamefont{Spergel}} \bibnamefont{and}
  \bibinfo{author}{\bibfnamefont{P.~J.} \bibnamefont{Steinhardt}},
  \bibinfo{journal}{Phys. Rev. Lett.} \textbf{\bibinfo{volume}{84}},
  \bibinfo{pages}{3760} (\bibinfo{year}{2000}), \eprint{astro-ph/9909386}.

\bibitem[{\citenamefont{Firmani et~al.}(2000)\citenamefont{Firmani, D'Onghia,
  Avila-Reese, Chincarini, and Hernandez}}]{Firmani:2000ce}
\bibinfo{author}{\bibfnamefont{C.}~\bibnamefont{Firmani}},
  \bibinfo{author}{\bibfnamefont{E.}~\bibnamefont{D'Onghia}},
  \bibinfo{author}{\bibfnamefont{V.}~\bibnamefont{Avila-Reese}},
  \bibinfo{author}{\bibfnamefont{G.}~\bibnamefont{Chincarini}},
  \bibnamefont{and}
  \bibinfo{author}{\bibfnamefont{X.}~\bibnamefont{Hernandez}},
  \bibinfo{journal}{Mon. Not. Roy. Astron. Soc.}
  \textbf{\bibinfo{volume}{315}}, \bibinfo{pages}{L29} (\bibinfo{year}{2000}),
  \eprint{astro-ph/0002376}.

\bibitem[{\citenamefont{Sand et~al.}(2004)\citenamefont{Sand, Treu, Smith, and
  Ellis}}]{Sand:2003bp}
\bibinfo{author}{\bibfnamefont{D.~J.} \bibnamefont{Sand}},
  \bibinfo{author}{\bibfnamefont{T.}~\bibnamefont{Treu}},
  \bibinfo{author}{\bibfnamefont{G.~P.} \bibnamefont{Smith}}, \bibnamefont{and}
  \bibinfo{author}{\bibfnamefont{R.~S.} \bibnamefont{Ellis}},
  \bibinfo{journal}{Astrophys. J.} \textbf{\bibinfo{volume}{604}},
  \bibinfo{pages}{88} (\bibinfo{year}{2004}), \eprint{astro-ph/0309465}.

\bibitem[{\citenamefont{Newman et~al.}(2013)\citenamefont{Newman, Treu, Ellis,
  and Sand}}]{Newman:2012nw}
\bibinfo{author}{\bibfnamefont{A.~B.} \bibnamefont{Newman}},
  \bibinfo{author}{\bibfnamefont{T.}~\bibnamefont{Treu}},
  \bibinfo{author}{\bibfnamefont{R.~S.} \bibnamefont{Ellis}}, \bibnamefont{and}
  \bibinfo{author}{\bibfnamefont{D.~J.} \bibnamefont{Sand}},
  \bibinfo{journal}{Astrophys. J.} \textbf{\bibinfo{volume}{765}},
  \bibinfo{pages}{25} (\bibinfo{year}{2013}), \eprint{1209.1392}.

\bibitem[{\citenamefont{Flores and Primack}(1994)}]{Flores:1994gz}
\bibinfo{author}{\bibfnamefont{R.~A.} \bibnamefont{Flores}} \bibnamefont{and}
  \bibinfo{author}{\bibfnamefont{J.~R.} \bibnamefont{Primack}},
  \bibinfo{journal}{Astrophys. J.} \textbf{\bibinfo{volume}{427}},
  \bibinfo{pages}{L1} (\bibinfo{year}{1994}), \eprint{astro-ph/9402004}.

\bibitem[{\citenamefont{Walker and Penarrubia}(2011)}]{Walker:2011zu}
\bibinfo{author}{\bibfnamefont{M.~G.} \bibnamefont{Walker}} \bibnamefont{and}
  \bibinfo{author}{\bibfnamefont{J.}~\bibnamefont{Penarrubia}},
  \bibinfo{journal}{Astrophys. J.} \textbf{\bibinfo{volume}{742}},
  \bibinfo{pages}{20} (\bibinfo{year}{2011}), \eprint{1108.2404}.

\bibitem[{\citenamefont{de~Blok et~al.}(2001)\citenamefont{de~Blok, McGaugh,
  Bosma, and Rubin}}]{deBlok:2001hbg}
\bibinfo{author}{\bibfnamefont{W.~J.~G.} \bibnamefont{de~Blok}},
  \bibinfo{author}{\bibfnamefont{S.~S.} \bibnamefont{McGaugh}},
  \bibinfo{author}{\bibfnamefont{A.}~\bibnamefont{Bosma}}, \bibnamefont{and}
  \bibinfo{author}{\bibfnamefont{V.~C.} \bibnamefont{Rubin}},
  \bibinfo{journal}{Astrophys. J.} \textbf{\bibinfo{volume}{552}},
  \bibinfo{pages}{L23} (\bibinfo{year}{2001}), \eprint{astro-ph/0103102}.

\bibitem[{\citenamefont{de~Blok and Bosma}(2002)}]{deBlok:2002vgq}
\bibinfo{author}{\bibfnamefont{W.~J.~G.} \bibnamefont{de~Blok}}
  \bibnamefont{and} \bibinfo{author}{\bibfnamefont{A.}~\bibnamefont{Bosma}},
  \bibinfo{journal}{Astron. Astrophys.} \textbf{\bibinfo{volume}{385}},
  \bibinfo{pages}{816} (\bibinfo{year}{2002}), \eprint{astro-ph/0201276}.

\bibitem[{\citenamefont{Simon et~al.}(2005)\citenamefont{Simon, Bolatto, Leroy,
  Blitz, and Gates}}]{Simon:2004sr}
\bibinfo{author}{\bibfnamefont{J.~D.} \bibnamefont{Simon}},
  \bibinfo{author}{\bibfnamefont{A.~D.} \bibnamefont{Bolatto}},
  \bibinfo{author}{\bibfnamefont{A.}~\bibnamefont{Leroy}},
  \bibinfo{author}{\bibfnamefont{L.}~\bibnamefont{Blitz}}, \bibnamefont{and}
  \bibinfo{author}{\bibfnamefont{E.~L.} \bibnamefont{Gates}},
  \bibinfo{journal}{Astrophys. J.} \textbf{\bibinfo{volume}{621}},
  \bibinfo{pages}{757} (\bibinfo{year}{2005}), \eprint{astro-ph/0412035}.

\bibitem[{\citenamefont{Tulin and Yu}(2018)}]{Tulin:2017ara}
\bibinfo{author}{\bibfnamefont{S.}~\bibnamefont{Tulin}} \bibnamefont{and}
  \bibinfo{author}{\bibfnamefont{H.-B.} \bibnamefont{Yu}},
  \bibinfo{journal}{Phys. Rept.} \textbf{\bibinfo{volume}{730}},
  \bibinfo{pages}{1} (\bibinfo{year}{2018}), \eprint{1705.02358}.

\bibitem[{\citenamefont{Kaplinghat et~al.}(2016)\citenamefont{Kaplinghat,
  Tulin, and Yu}}]{Kaplinghat:2015aga}
\bibinfo{author}{\bibfnamefont{M.}~\bibnamefont{Kaplinghat}},
  \bibinfo{author}{\bibfnamefont{S.}~\bibnamefont{Tulin}}, \bibnamefont{and}
  \bibinfo{author}{\bibfnamefont{H.-B.} \bibnamefont{Yu}},
  \bibinfo{journal}{Phys. Rev. Lett.} \textbf{\bibinfo{volume}{116}},
  \bibinfo{pages}{041302} (\bibinfo{year}{2016}), \eprint{1508.03339}.

\bibitem[{\citenamefont{McDermott}(2018)}]{McDermott:2017vyk}
\bibinfo{author}{\bibfnamefont{S.~D.} \bibnamefont{McDermott}},
  \bibinfo{journal}{Phys. Rev. Lett.} \textbf{\bibinfo{volume}{120}},
  \bibinfo{pages}{221806} (\bibinfo{year}{2018}), \eprint{1711.00857}.

\bibitem[{\citenamefont{Loeb and Weiner}(2011)}]{Loeb:2010gj}
\bibinfo{author}{\bibfnamefont{A.}~\bibnamefont{Loeb}} \bibnamefont{and}
  \bibinfo{author}{\bibfnamefont{N.}~\bibnamefont{Weiner}},
  \bibinfo{journal}{Phys. Rev. Lett.} \textbf{\bibinfo{volume}{106}},
  \bibinfo{pages}{171302} (\bibinfo{year}{2011}), \eprint{1011.6374}.

\bibitem[{\citenamefont{Schutz and Slatyer}(2015)}]{Schutz:2014nka}
\bibinfo{author}{\bibfnamefont{K.}~\bibnamefont{Schutz}} \bibnamefont{and}
  \bibinfo{author}{\bibfnamefont{T.~R.} \bibnamefont{Slatyer}},
  \bibinfo{journal}{JCAP} \textbf{\bibinfo{volume}{1501}}, \bibinfo{pages}{021}
  (\bibinfo{year}{2015}), \eprint{1409.2867}.

\bibitem[{\citenamefont{Chu and Garcia-Cely}(2018)}]{Chu:2018nki}
\bibinfo{author}{\bibfnamefont{X.}~\bibnamefont{Chu}} \bibnamefont{and}
  \bibinfo{author}{\bibfnamefont{C.}~\bibnamefont{Garcia-Cely}},
  \bibinfo{journal}{JCAP} \textbf{\bibinfo{volume}{1807}}, \bibinfo{pages}{013}
  (\bibinfo{year}{2018}), \eprint{1803.09762}.

\bibitem[{\citenamefont{Vogelsberger et~al.}(2018)\citenamefont{Vogelsberger,
  Zavala, Schutz, and Slatyer}}]{Vogelsberger:2018bok}
\bibinfo{author}{\bibfnamefont{M.}~\bibnamefont{Vogelsberger}},
  \bibinfo{author}{\bibfnamefont{J.}~\bibnamefont{Zavala}},
  \bibinfo{author}{\bibfnamefont{K.}~\bibnamefont{Schutz}}, \bibnamefont{and}
  \bibinfo{author}{\bibfnamefont{T.~R.} \bibnamefont{Slatyer}}
  (\bibinfo{year}{2018}), \eprint{1805.03203}.

\bibitem[{Not({\natexlab{a}})}]{Note1}
\bibinfo{note}{See Supplemental Material for the details on: 1) the assignment
  of halo formation time $t_\ast$; 2) the implementation of heat conduction
  from DM self-interaction; 3) the justification of the assumption of
  quasi-static equilibrium; 4) the initial condition dependence of our
  analysis; 5) and the choice of the fudge factor $b=3$.}

\bibitem[{\citenamefont{Dutton and Maccio}(2014)}]{Dutton:2014xda}
\bibinfo{author}{\bibfnamefont{A.~A.} \bibnamefont{Dutton}} \bibnamefont{and}
  \bibinfo{author}{\bibfnamefont{A.~V.} \bibnamefont{Maccio}},
  \bibinfo{journal}{Mon. Not. Roy. Astron. Soc.}
  \textbf{\bibinfo{volume}{441}}, \bibinfo{pages}{3359} (\bibinfo{year}{2014}),
  \eprint{1402.7073}.

\bibitem[{\citenamefont{Lynden-Bell and Eggleton}(1980)}]{1980MNRAS.191..483L}
\bibinfo{author}{\bibfnamefont{D.}~\bibnamefont{Lynden-Bell}} \bibnamefont{and}
  \bibinfo{author}{\bibfnamefont{P.~P.} \bibnamefont{Eggleton}},
  \bibinfo{journal}{Mon. Not. Roy. Astron. Soc.}
  \textbf{\bibinfo{volume}{191}}, \bibinfo{pages}{483} (\bibinfo{year}{1980}).

\bibitem[{\citenamefont{Binney and Tremaine}(2011)}]{GalacticDynamics}
\bibinfo{author}{\bibfnamefont{J.}~\bibnamefont{Binney}} \bibnamefont{and}
  \bibinfo{author}{\bibfnamefont{S.}~\bibnamefont{Tremaine}},
  \emph{\bibinfo{title}{Galactic Dynamics: Second Edition}}, Princeton Series
  in Astrophysics (\bibinfo{publisher}{Princeton University Press},
  \bibinfo{year}{2011}).

\bibitem[{\citenamefont{Kamada et~al.}(2018{\natexlab{a}})\citenamefont{Kamada,
  Kim, Kim, and Sekiguchi}}]{Kamada:2017gfc}
\bibinfo{author}{\bibfnamefont{A.}~\bibnamefont{Kamada}},
  \bibinfo{author}{\bibfnamefont{H.~J.} \bibnamefont{Kim}},
  \bibinfo{author}{\bibfnamefont{H.}~\bibnamefont{Kim}}, \bibnamefont{and}
  \bibinfo{author}{\bibfnamefont{T.}~\bibnamefont{Sekiguchi}},
  \bibinfo{journal}{Phys. Rev. Lett.} \textbf{\bibinfo{volume}{120}},
  \bibinfo{pages}{131802} (\bibinfo{year}{2018}{\natexlab{a}}),
  \eprint{1707.09238}.

\bibitem[{\citenamefont{Kamada et~al.}(2018{\natexlab{b}})\citenamefont{Kamada,
  Kim, and Kim}}]{Kamada:2018hte}
\bibinfo{author}{\bibfnamefont{A.}~\bibnamefont{Kamada}},
  \bibinfo{author}{\bibfnamefont{H.~J.} \bibnamefont{Kim}}, \bibnamefont{and}
  \bibinfo{author}{\bibfnamefont{H.}~\bibnamefont{Kim}},
  \bibinfo{journal}{Phys. Rev.} \textbf{\bibinfo{volume}{D98}},
  \bibinfo{pages}{023509} (\bibinfo{year}{2018}{\natexlab{b}}),
  \eprint{1805.05648}.

\bibitem[{\citenamefont{Gilmore et~al.}(2007)\citenamefont{Gilmore, Wilkinson,
  Wyse, Kleyna, Koch, Evans, and Grebel}}]{Gilmore:2007fy}
\bibinfo{author}{\bibfnamefont{G.}~\bibnamefont{Gilmore}},
  \bibinfo{author}{\bibfnamefont{M.~I.} \bibnamefont{Wilkinson}},
  \bibinfo{author}{\bibfnamefont{R.~F.~G.} \bibnamefont{Wyse}},
  \bibinfo{author}{\bibfnamefont{J.~T.} \bibnamefont{Kleyna}},
  \bibinfo{author}{\bibfnamefont{A.}~\bibnamefont{Koch}},
  \bibinfo{author}{\bibfnamefont{N.~W.} \bibnamefont{Evans}}, \bibnamefont{and}
  \bibinfo{author}{\bibfnamefont{E.~K.} \bibnamefont{Grebel}},
  \bibinfo{journal}{Astrophys. J.} \textbf{\bibinfo{volume}{663}},
  \bibinfo{pages}{948} (\bibinfo{year}{2007}), \eprint{astro-ph/0703308}.

\bibitem[{\citenamefont{Balberg and Shapiro}(2002)}]{Balberg:2001qg}
\bibinfo{author}{\bibfnamefont{S.}~\bibnamefont{Balberg}} \bibnamefont{and}
  \bibinfo{author}{\bibfnamefont{S.~L.} \bibnamefont{Shapiro}},
  \bibinfo{journal}{Phys. Rev. Lett.} \textbf{\bibinfo{volume}{88}},
  \bibinfo{pages}{101301} (\bibinfo{year}{2002}), \eprint{astro-ph/0111176}.

\bibitem[{\citenamefont{Ahn and Shapiro}(2005)}]{Ahn:2004xt}
\bibinfo{author}{\bibfnamefont{K.-J.} \bibnamefont{Ahn}} \bibnamefont{and}
  \bibinfo{author}{\bibfnamefont{P.~R.} \bibnamefont{Shapiro}},
  \bibinfo{journal}{Mon. Not. Roy. Astron. Soc.}
  \textbf{\bibinfo{volume}{363}}, \bibinfo{pages}{1092} (\bibinfo{year}{2005}),
  \eprint{astro-ph/0412169}.

\bibitem[{\citenamefont{Koda and Shapiro}(2011)}]{Koda:2011yb}
\bibinfo{author}{\bibfnamefont{J.}~\bibnamefont{Koda}} \bibnamefont{and}
  \bibinfo{author}{\bibfnamefont{P.~R.} \bibnamefont{Shapiro}},
  \bibinfo{journal}{Mon. Not. Roy. Astron. Soc.}
  \textbf{\bibinfo{volume}{415}}, \bibinfo{pages}{1125} (\bibinfo{year}{2011}),
  \eprint{1101.3097}.

\bibitem[{\citenamefont{Pollack et~al.}(2015)\citenamefont{Pollack, Spergel,
  and Steinhardt}}]{Pollack:2014rja}
\bibinfo{author}{\bibfnamefont{J.}~\bibnamefont{Pollack}},
  \bibinfo{author}{\bibfnamefont{D.~N.} \bibnamefont{Spergel}},
  \bibnamefont{and} \bibinfo{author}{\bibfnamefont{P.~J.}
  \bibnamefont{Steinhardt}}, \bibinfo{journal}{Astrophys. J.}
  \textbf{\bibinfo{volume}{804}}, \bibinfo{pages}{131} (\bibinfo{year}{2015}),
  \eprint{1501.00017}.

\bibitem[{\citenamefont{Essig et~al.}(2018)\citenamefont{Essig, Yu, Zhong, and
  Mcdermott}}]{Essig:2018pzq}
\bibinfo{author}{\bibfnamefont{R.}~\bibnamefont{Essig}},
  \bibinfo{author}{\bibfnamefont{H.-B.} \bibnamefont{Yu}},
  \bibinfo{author}{\bibfnamefont{Y.-M.} \bibnamefont{Zhong}}, \bibnamefont{and}
  \bibinfo{author}{\bibfnamefont{S.~D.} \bibnamefont{Mcdermott}}
  (\bibinfo{year}{2018}), \eprint{1809.01144}.

\bibitem[{\citenamefont{Nishikawa et~al.}(2019)\citenamefont{Nishikawa, Boddy,
  and Kaplinghat}}]{Nishikawa:2019lsc}
\bibinfo{author}{\bibfnamefont{H.}~\bibnamefont{Nishikawa}},
  \bibinfo{author}{\bibfnamefont{K.~K.} \bibnamefont{Boddy}}, \bibnamefont{and}
  \bibinfo{author}{\bibfnamefont{M.}~\bibnamefont{Kaplinghat}}
  (\bibinfo{year}{2019}), \eprint{1901.00499}.

\bibitem[{\citenamefont{Navarro et~al.}(1996)\citenamefont{Navarro, Frenk, and
  White}}]{Navarro:1995iw}
\bibinfo{author}{\bibfnamefont{J.~F.} \bibnamefont{Navarro}},
  \bibinfo{author}{\bibfnamefont{C.~S.} \bibnamefont{Frenk}}, \bibnamefont{and}
  \bibinfo{author}{\bibfnamefont{S.~D.~M.} \bibnamefont{White}},
  \bibinfo{journal}{Astrophys. J.} \textbf{\bibinfo{volume}{462}},
  \bibinfo{pages}{563} (\bibinfo{year}{1996}), \eprint{astro-ph/9508025}.

\bibitem[{\citenamefont{Navarro et~al.}(1997)\citenamefont{Navarro, Frenk, and
  White}}]{Navarro:1996gj}
\bibinfo{author}{\bibfnamefont{J.~F.} \bibnamefont{Navarro}},
  \bibinfo{author}{\bibfnamefont{C.~S.} \bibnamefont{Frenk}}, \bibnamefont{and}
  \bibinfo{author}{\bibfnamefont{S.~D.~M.} \bibnamefont{White}},
  \bibinfo{journal}{Astrophys. J.} \textbf{\bibinfo{volume}{490}},
  \bibinfo{pages}{493} (\bibinfo{year}{1997}), \eprint{astro-ph/9611107}.

\bibitem[{\citenamefont{Balberg et~al.}(2002)\citenamefont{Balberg, Shapiro,
  and Inagaki}}]{Balberg:2002ue}
\bibinfo{author}{\bibfnamefont{S.}~\bibnamefont{Balberg}},
  \bibinfo{author}{\bibfnamefont{S.~L.} \bibnamefont{Shapiro}},
  \bibnamefont{and} \bibinfo{author}{\bibfnamefont{S.}~\bibnamefont{Inagaki}},
  \bibinfo{journal}{Astrophys. J.} \textbf{\bibinfo{volume}{568}},
  \bibinfo{pages}{475} (\bibinfo{year}{2002}), \eprint{astro-ph/0110561}.

\bibitem[{Not({\natexlab{b}})}]{Note2}
\bibinfo{note}{For the SIDM cross section in our analysis ($\sigma_{\rm
  self}/m\sim{\cal O}(1)\,{\rm cm^2/g}$), the halo evolution takes place in the
  long-mean-free-path (LMFP) regime; the DM free-streaming length $\lambda$ is
  much larger than the system size (Jean's length) $H=\sqrt{\nu^2/4\pi G\rho}$.
  In this regime, the RHS of Eq.~\eqref{eq:1entropy} is dominated by the
  effects of self-heating ($\delta u_{\rm semi}\delta t>\delta u_{\rm
  cond}/\delta t$) for $t_{\rm heat}\lesssim t_{\rm self}$. We numerically
  confirm that the estimation given in Eq.~\eqref{eq:denestimate} works well
  for any halo mass and for $\sigma_{\rm self}/m\lesssim 10\,{\rm cm}^2/{\rm
  g}$, as long as $t_{\rm heat}\lesssim t_{\rm self}(r=r_s)$. This estimate
  resembles the analytic estimation given in Ref.~\cite{Chu:2018nki}, although
  Ref.~\cite{Chu:2018nki} assumes a uniform and constant velocity dispersion
  profile.}

\bibitem[{\citenamefont{Wolf et~al.}(2010)\citenamefont{Wolf, Martinez,
  Bullock, Kaplinghat, Geha, Munoz, Simon, and Avedo}}]{Wolf:2009tu}
\bibinfo{author}{\bibfnamefont{J.}~\bibnamefont{Wolf}},
  \bibinfo{author}{\bibfnamefont{G.~D.} \bibnamefont{Martinez}},
  \bibinfo{author}{\bibfnamefont{J.~S.} \bibnamefont{Bullock}},
  \bibinfo{author}{\bibfnamefont{M.}~\bibnamefont{Kaplinghat}},
  \bibinfo{author}{\bibfnamefont{M.}~\bibnamefont{Geha}},
  \bibinfo{author}{\bibfnamefont{R.~R.} \bibnamefont{Munoz}},
  \bibinfo{author}{\bibfnamefont{J.~D.} \bibnamefont{Simon}}, \bibnamefont{and}
  \bibinfo{author}{\bibfnamefont{F.~F.} \bibnamefont{Avedo}},
  \bibinfo{journal}{Mon. Not. Roy. Astron. Soc.}
  \textbf{\bibinfo{volume}{406}}, \bibinfo{pages}{1220} (\bibinfo{year}{2010}),
  \eprint{0908.2995}.

\bibitem[{\citenamefont{Press and Schechter}(1974)}]{Press:1973iz}
\bibinfo{author}{\bibfnamefont{W.~H.} \bibnamefont{Press}} \bibnamefont{and}
  \bibinfo{author}{\bibfnamefont{P.}~\bibnamefont{Schechter}},
  \bibinfo{journal}{Astrophys. J.} \textbf{\bibinfo{volume}{187}},
  \bibinfo{pages}{425} (\bibinfo{year}{1974}).

\bibitem[{\citenamefont{Bardeen et~al.}(1986)\citenamefont{Bardeen, Bond,
  Kaiser, and Szalay}}]{Bardeen:1985tr}
\bibinfo{author}{\bibfnamefont{J.~M.} \bibnamefont{Bardeen}},
  \bibinfo{author}{\bibfnamefont{J.~R.} \bibnamefont{Bond}},
  \bibinfo{author}{\bibfnamefont{N.}~\bibnamefont{Kaiser}}, \bibnamefont{and}
  \bibinfo{author}{\bibfnamefont{A.~S.} \bibnamefont{Szalay}},
  \bibinfo{journal}{Astrophys. J.} \textbf{\bibinfo{volume}{304}},
  \bibinfo{pages}{15} (\bibinfo{year}{1986}).

\bibitem[{\citenamefont{Cheng et~al.}(2015)\citenamefont{Cheng, Chu, and
  Tang}}]{Cheng:2015dga}
\bibinfo{author}{\bibfnamefont{D.}~\bibnamefont{Cheng}},
  \bibinfo{author}{\bibfnamefont{M.~C.} \bibnamefont{Chu}}, \bibnamefont{and}
  \bibinfo{author}{\bibfnamefont{J.}~\bibnamefont{Tang}},
  \bibinfo{journal}{JCAP} \textbf{\bibinfo{volume}{1507}}, \bibinfo{pages}{009}
  (\bibinfo{year}{2015}), \eprint{1503.05682}.

\bibitem[{\citenamefont{Pitaevskii et~al.}(1981)\citenamefont{Pitaevskii,
  Lifshitz, and Skykes}}]{Lifshitz:1981}
\bibinfo{author}{\bibfnamefont{L.}~\bibnamefont{Pitaevskii}},
  \bibinfo{author}{\bibfnamefont{E.}~\bibnamefont{Lifshitz}}, \bibnamefont{and}
  \bibinfo{author}{\bibfnamefont{J.}~\bibnamefont{Skykes}},
  \emph{\bibinfo{title}{Physical Kinetics}}, Course of Theoretical Physics S
  (\bibinfo{publisher}{Pergamon Press}, \bibinfo{year}{1981}).

\bibitem[{\citenamefont{Chapman et~al.}(1990)\citenamefont{Chapman, Cowling,
  Burnett, and Cercignani}}]{Chapman:1990}
\bibinfo{author}{\bibfnamefont{S.}~\bibnamefont{Chapman}},
  \bibinfo{author}{\bibfnamefont{T.}~\bibnamefont{Cowling}},
  \bibinfo{author}{\bibfnamefont{D.}~\bibnamefont{Burnett}}, \bibnamefont{and}
  \bibinfo{author}{\bibfnamefont{C.}~\bibnamefont{Cercignani}},
  \emph{\bibinfo{title}{The Mathematical Theory of Non-uniform Gases: An
  Account of the Kinetic Theory of Viscosity, Thermal Conduction and Diffusion
  in Gases}}, Cambridge Mathematical Library (\bibinfo{publisher}{Cambridge
  University Press}, \bibinfo{year}{1990}).

\end{thebibliography}
\newpage
\appendix
\section{Supplemental Material}

In this supplemental material, we provide details on the assignment of halo formation time $t_\ast$, the implementation of heat conduction from DM self-interaction, the justification of the assumption of quasi-static equilibrium, the initial condition dependence of our analysis, and the choice of the fudge factor $b=3$.
\\
\paragraph{\underline{Halo formation time}.---}

The structure formation is hierarchical in the sense that smaller structures form prior to larger ones.
Since we consider evolution of halos in a wide mass range ($M_{200}=10^{8}$--$10^{15}\,{\rm M}_\odot$), we take into account the formation time of halos consistent with their mass.
In this section, we summarize the assignment of the halo formation time $t_\ast$.

We define the overdensity $\delta=\rho_m/\bar{\rho}_m-1$, where $\bar{\rho}_m$ is the average matter density.
For a standard CDM model, the overdensity in the linear regime ($\delta<1$) grows as $\delta(t)=\delta(t_i)D(t)/D(t_i)$, where $D(t)$ is the growth factor given by
\begin{equation}
D\left(a(t)\right)=\frac{H(a)\int^a_0 da^{\prime}/\left[a^\prime H(a^\prime)/H_0\right]^3}{H_0\int^1_0 da^{\prime}/\left[a^\prime H(a^\prime)/H_0\right]^3}\,,
\label{eq:CDMgf}
\end{equation}
which is normalized as $D\left(a(t_{\rm age})\right)=1$.
According to the spherical collapse model~\cite{GalacticDynamics} in the flat matter dominated FRW Universe, virialized halos form when their overdensity becomes $\delta(t_\ast)\sim200$, where the linear theory is no longer valid.
On the other hand, if we assume that the linear theory were valid even when $\delta>1$, a virialized halo is formed when the overdensity in the linear theory exceeds some critical value $\delta_{\rm c}\simeq 1.686$ at time $t_\ast$: $\delta_L(t_\ast)\simeq\delta_{\rm c}$.

\begin{figure}
\centering
\includegraphics[scale=0.55]{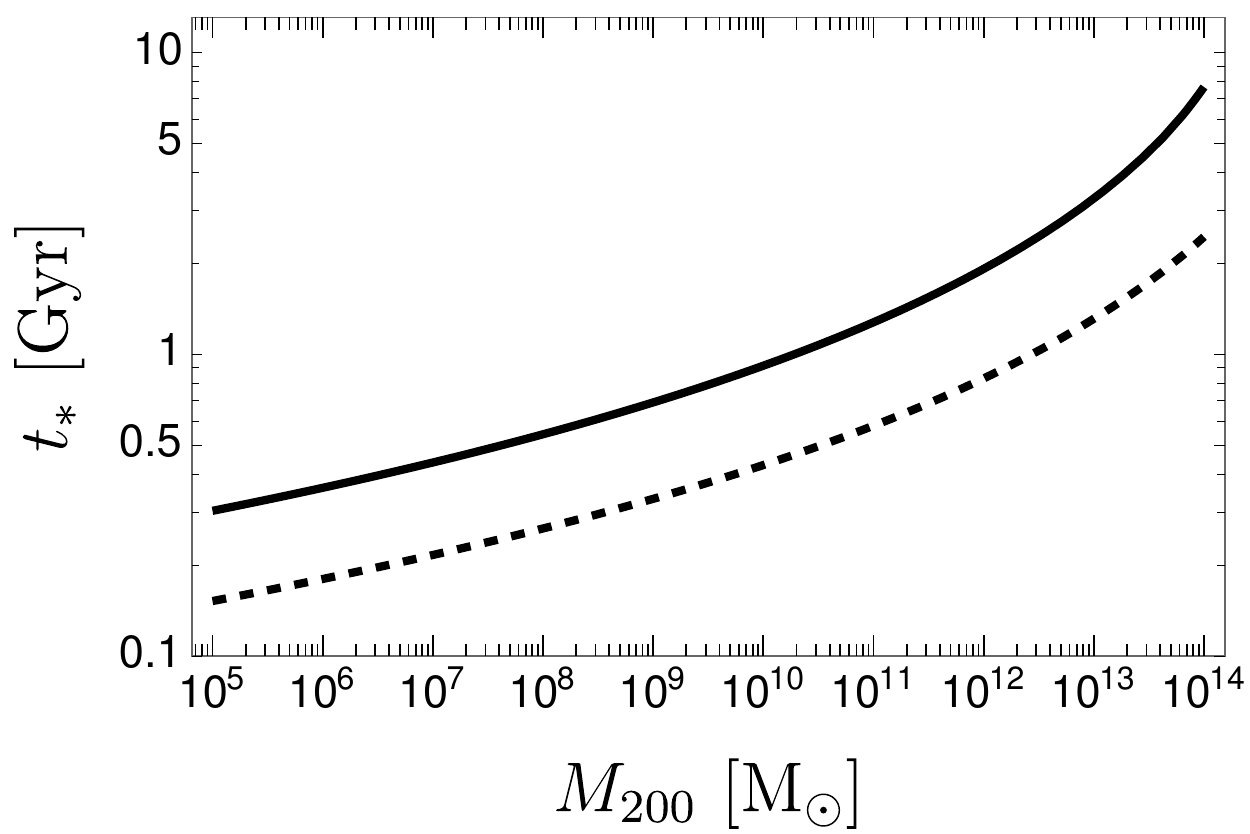}
\caption{
The halo-mass dependence of the halo formation time $t_\ast$.
The assignment of the formation time is based on the spherical collapse model, and the observed CDM density fluctuations.
The solid curve corresponds to the condition $2\sigma_M(t_\ast)=\delta_{\rm c}$.
The dashed curve is the one that we use in this work, which is determined by the condition $3\sigma_M(t_\ast)=\delta_{\rm c}$.
}
\label{fig:tformation}
\end{figure}

Now we would like to assign masses to such virialized halos.
To do so, we define a smoothed overdensity $\delta_M$, which is the overdensity after averaging out all density fluctuations on scales less than scale $R$:
\begin{equation}
\delta_M(\boldsymbol{x})=\int d^3 \boldsymbol{x}^\prime W_R(\boldsymbol{x}^\prime-\boldsymbol{x})\delta_L(\boldsymbol{x}^\prime)\,,
\end{equation}
where we choose the window function as the top-hat filter in real space; $W_R(\boldsymbol{x}^\prime-\boldsymbol{x})=3/(4\pi R^3) \Theta(R-|\boldsymbol{x}^\prime-\boldsymbol{x}|)$.
Given a length scale $R$, we assign a mass scale of the filter as $M=4\pi R^3\bar{\rho}_m/3$.
Following the Press-Schechter formalism~\cite{Press:1973iz}, we identify the mass fraction that is contained in halos of mass greater than $M$ with the probability that $\delta_M>\delta_{\rm c}$;
\begin{equation}
\begin{aligned}
p\left(\delta_M>\delta_{\rm c}\right)=\int^{\infty}_{\delta_{\rm c}}\frac{\exp\left(-\frac{\delta_M^2}{2\sigma_M^2}\right)}{\sqrt{2\pi}\sigma_M}d\delta_M \,,
\end{aligned}
\label{eq:PSprob}
\end{equation}
where $\sigma_M^2$ is the variance of $\delta_M$ given by
\begin{equation}
\begin{aligned}
\sigma_M^2(t)=D^2\left(a(t)\right)\int^{\infty}_{0}\frac{k^2 dk}{2\pi^2}P_L(k)\widetilde{W}^2_R(k)\,,
\end{aligned}
\label{eq:CDMPL}
\end{equation}
where $P_L(k)$ is the CDM linear matter power spectrum at present, and $\widetilde{W}_R(k)$ is the Fourier transformed top-hat window function.
Here, we take $P_L(k)\propto T_{\rm BBKS}^2(k)P_{\rm prim}(k)$;
$T_{\rm BBKS}(k)$ is the BBKS fitting transfer function~\cite{Bardeen:1985tr} and $P_{\rm prim}(k)\propto k^{n_s}$.
We adopt the following cosmological parameters as in Ref.~\cite{Cheng:2015dga}: $\Omega_m=0.3$, $\Omega_{\Lambda}=0.7$, $n_s=0.96$, and $\sigma_8=0.8$.
The overall normalization for $P_L(k)$ is set by the cosmological parameter $\sigma_8$.
From Eq.~\eqref{eq:PSprob}, the fraction of mass contained in halos with mass greater than $M$ becomes significant when $\sigma_M(t)\sim\delta_{\rm c}$.
We choose the formation time of a halo of mass $M$ to be the moment when $3\sigma_M(t_\ast)=\delta_{\rm c}$, where the factor $3$ is chosen to reproduce the halo evolution time used in Ref.~\cite{Kaplinghat:2015aga}.
The halo mass dependence of $t_\ast$ is shown in FIG.~\ref{fig:tformation}.
\\
\paragraph{\underline{Heat conduction from self-interaction of DM}.---}
The elastic self-interaction of DM particles allow for a fluid element to conduct heat with neighboring sites in the radial direction.
We model the heat conduction effect with a heat diffusion equation:
\begin{equation}
\frac{L}{4\pi r^{2}}=-\kappa m \nabla \nu^2\,,
\label{eq:Fourier}
\end{equation}
where $L(r,t)$ is the luminosity, that is defined to be the rate of energy that crosses a sphere at radius $r$ in the outward radial direction.
$\kappa$ is the thermal conductivity.
It is very challenging to determine the general expression of $\kappa$ in terms of $\sigma_{\rm self}/m$ from first principles.
If the free-streaming length $\lambda= 1/(\rho \sigma_{\rm self}/m)$ is much shorter than the system size (Jean's length) $H=\sqrt{\nu^2/4\pi G\rho}$, the Fourier law of heat flux, Eq.~\eqref{eq:Fourier}, can be derived from the Boltzmann equation~\cite{Lifshitz:1981,Chapman:1990}.
In this short mean free path (SMFP) regime ($\lambda\ll H$), $\kappa$ is expressed as 
\begin{equation}
\kappa_{\rm SMFP}=\frac{75\sqrt{\pi}}{64}\frac{\rho\lambda^2}{am t_{\rm self}}\,,
\end{equation}
where $t_{\rm self}=1/(a\rho\nu\sigma_{\rm self}/m)$ is the self-interaction time scale with $a=\sqrt{16/\pi}$.
In the long mean free path (LMFP) regime ($\lambda\gg H$), there is an empirical thermal conduction formula similar to the one in the SMFP regime~\cite{1980MNRAS.191..483L}:
\begin{equation}
\kappa_{\rm LMFP}=\frac{3C}{2}\frac{\rho H^2}{mt_{\rm self}}\,,
\end{equation}
where $C\simeq 0.75$ is a constant that cannot be determined analytically, and it is calibrated with $N$-body simulations in the LMFP regime~\cite{Balberg:2002ue,Koda:2011yb}.
We naively interpolate the thermal conductivity in the two regimes as $\kappa^{-1}=\kappa^{-1}_{\rm SMFP}+\kappa^{-1}_{\rm LMFP}$.
The rate of specific heat gain for a fluid element is given as
\begin{equation}
\frac{\delta u_{\rm cond}}{\delta t}=-\frac{1}{4\pi r^2\rho}\frac{\partial L}{\partial r}\,.
\end{equation}
\\
\paragraph{\underline{Assumption of quasi-static hydrostatic equilibrium}.---}
\begin{figure*}
\centering
\includegraphics[scale=0.5]{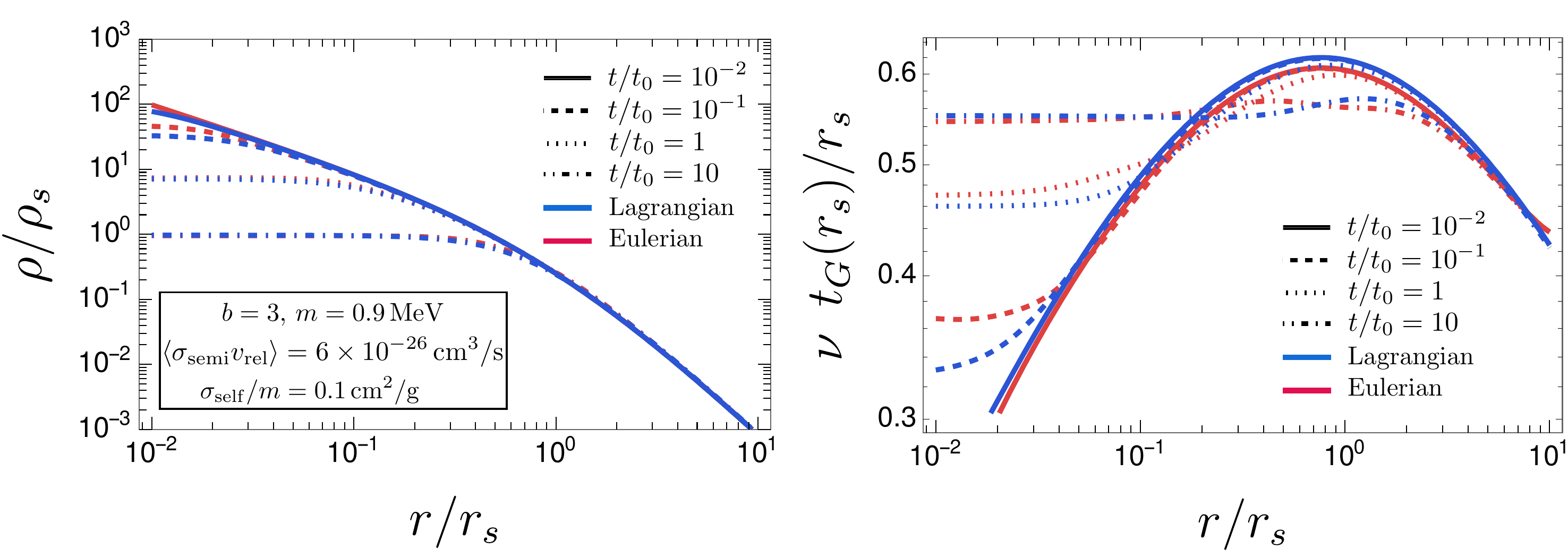}
\caption{
{\it Left panel} - The time evolution of DM density $\rho$ of a MW satellite galaxy ($M_{200}=10^9\,{\rm M}_\odot$).
NFW profile ($\rho_s=0.02\,M_{\odot}/{\rm pc^{3}}$, $r_s=1.26\,\kpc$) is used as the initial condition.
Heat conduction effect through self-interaction of DM is not taken into account.
Numerical solutions from the Lagrangian formulation [Eqs.~\eqref{eq:gravothermaleqns}] and the Eulerian formulation [Eqs.~\eqref{eq:Eulerfluid}] are in blue and red, respectively.
The reference time scale is $t_0=1\,{\rm Gyr}$. 
The set the fudge factor $b=3$ (see Eq.~\eqref{eq:xi}), and the cross sections are $\left\langle \sigma_{\rm semi}v_{\rm rel}\right\rangle/m=6\times10^{-26}\,{\rm cm^3/s}$ and $\sigma_{\rm self}/m=0.1\,{\rm cm^2/g}$ with SHDM mass $m=0.9\,{\rm MeV}$.
We find that two formalisms results in consistent profiles for $\rho$.
The agreement between two formalisms validates the quasi-static equilibrium assumption in the Lagrangian formulation.
{\it Right panel} -Same as the left panel but for the velocity dispersion $\nu$.
$t_{G}(r_s)=1/\sqrt{4\pi G\rho_{\rm NFW}(r_s)}$ is the gravitational time scale with $\rho_{\rm NFW}(r_s)=\rho_s/4$.
}
\label{fig:formalisms}
\end{figure*}
In the adopted gravothermal model [Eqs.~\eqref{eq:gravothermaleqns}], we assume that the evolution of a halo is quasi-static:
we require that the hydrostatic equilibrium condition, Eq.~\eqref{eq:1Euler}, is held at each moment.
This may be a good approximation as long as the gravitational time scale, $t_G$, is the shortest time scale of the system~\cite{Balberg:2001qg,Balberg:2002ue}:
\begin{equation}
\begin{aligned}
t_{G}&=\left(4\pi G\rho\right)^{-1/2}\\
&\simeq7.8\times10^7\,{\rm yr}\left(\frac{0.011\,M_{\odot}/\pc^{3}}{\rho_s}\right)^{1/2}\,,
\end{aligned}
\label{eq:tG}
\end{equation}
where in the second equality, we took $\rho=\rho_{\rm NFW}(r_s)=\rho_s/4$ as the representative value for a halo.
In the case of pure SIDM, the approximation is confirmed to be in good agreement with the $N$-body simulations~\cite{Koda:2011yb,Nishikawa:2019lsc}.

Since we are interested in the case where the two other time scales, $t_{\rm heat}$ and $t_{\rm self}$, are comparable to the age of the Universe, the same approximation is expected to be valid.
Nevertheless, it is good to have an independent check for the case of SHDM, since we take into account the self-heating from the semi-annihilation of DM.
The comparison between our fluid model and the $N$-body simulations would be the most ideal check,
but performing $N$-body simulations for SHDM is beyond the scope of this work.
Instead, we compare our results to the solutions of the gravothermal fluid model in the Eulerian description of the DM fluid, where the quasi-static equilibrium is not assumed~\cite{Ahn:2004xt,Chu:2018nki}:
\begin{subequations}
\begin{align}
&\frac{\partial\rho}{\partial t}+\frac{1}{r^2}\frac{\partial}{\partial r}\left(r^{2}\rho V_{r}\right)=0\label{eq:Eulera}\,,\\
&\frac{\partial V_{r}}{\partial t}+V_{r}\frac{\partial V_{r}}{\partial r}=-\frac{\partial\Phi}{\partial r}-\frac{1}{\rho}\frac{\partial\left(\rho\nu^{2}\right)}{\partial r}\label{eq:Eulerb}\,,\\
&\frac{1}{r^{2}}\frac{\partial}{\partial r}\left(r^{2}\frac{\partial\Phi}{\partial r}\right)=4\pi G \rho\label{eq:Eulerc}\,,\\
&\frac{1}{r^{2}}\frac{\partial}{\partial r}\left(r^{2}V_{r}\right)+\frac{3}{\nu}\left[\frac{\partial\nu}{\partial t}+V_{r}\frac{\partial\nu}{\partial r}\right]=\frac{1}{\nu^{2}}\frac{\delta u}{\delta t}\label{eq:Eulerd}\,,
\end{align}
\label{eq:Eulerfluid}
\end{subequations}
where $V_r$ is the fluid bulk velocity in the radial direction, and $\Phi$ is the gravitational potential.
Remark that quasi-static equilibrium corresponds to taking $V_r=0$ in Eq.~\eqref{eq:Eulerb}, which becomes Eq.~\eqref{eq:1Euler} in this limit.
We solved Eqs.~\eqref{eq:Eulerfluid} iteratively, which is summarized into three procedures:
1) assume a uniform and constant $\nu$ and integrate Eqs.~\eqref{eq:Eulera} and \eqref{eq:Eulerd} to get solutions for $\rho$ and $V_r$;
2) take $\rho$ and $V_r$ from the first procedure and integrate Eqs.~\eqref{eq:Eulerb} and \eqref{eq:Eulerc} to get a solution for $\nu$;
and 3) repeat the first procedure but with the updated $\nu$ from the second procedure.
We repeat this until the solutions converge.

Due to numerical difficulties, we only compare the solutions in the case where heat conduction effect is absent;
we take $\delta u=\delta u_{\rm semi}$ (see Eq.~\eqref{eq:semiinjection}) in both Eq.~\eqref{eq:1entropy} and Eq.~\eqref{eq:Eulerd} for the comparison.
The time evolution of $\rho$ and $\nu$ in both formalisms is presented in FIG.~\ref{fig:formalisms}.
We find that both Eqs.~\eqref{eq:gravothermaleqns} (in blue) and Eqs.~\eqref{eq:Eulerfluid} (in red) exhibit consistent resultant profiles for $\rho$ and $\nu$.
Since the quasi-static equilibrium is not assumed in Eqs.~\eqref{eq:Eulerfluid}, we conclude that dynamics induced by heat injection from semi-annihilation do not spoil the assumption.
\\
\paragraph{\underline{Cored initial halo profiles}.---}
\begin{figure*}
\centering
\includegraphics[scale=0.55]{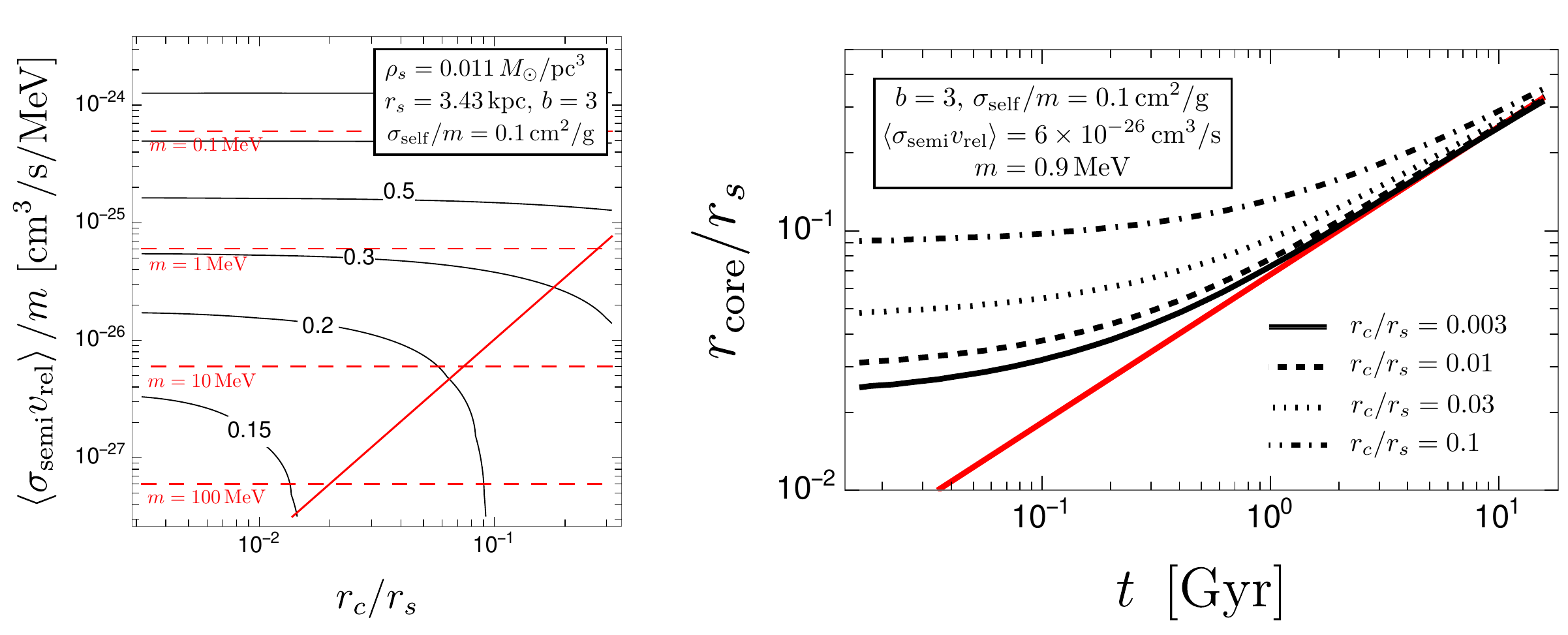}
\caption{
{\it Left panel} - Contour plot for the resultant core radius ($r_{\rm core}/r_s$) of a field dwarf/LSB galaxy ($M_{200}=10^{10}\,{\rm M}_\odot$).
The horizontal axis ($r_c/r_s$) is the core size of the initial CNFW profile.
We set the fudge factor $b=3$ (see Eq.~\eqref{eq:xi}).
The red dashed lines indicate the canonical cross section $\left\langle\sigma_{\rm semi}v_{\rm rel}\right\rangle=6\times10^{-26}\,{\rm cm^3/s}$ for different SHDM masses.
The red solid line is the contour where $r_c$ is identical to the analytic estimate for the resultant core radius in the $r_c=0$ case (see Eq.~\eqref{eq:restimate}).
We find that our benchmark parameter space ($\left\langle\sigma_{\rm semi}v_{\rm rel}\right\rangle\simeq6\times10^{-26}\,{\rm cm^3/s}$, $\sigma_{\rm self}\simeq0.1\,{\rm cm^2/g}$, $m=0.1$--$10\,{\rm MeV}$) is insensitive to the initial CNFW profile as long as $r_c/r_s\lesssim0.1$.
{\it Right panel} -Time evolution of the core radius of the same field dwarf/LSB halo in the left panel.
The black lines correspond to different values of $r_c$.
The red solid line is the analytic estimate for the core radius in the $r_c=0$ case (given in Eq.~\eqref{eq:restimate}).
All the black lines approach the red line in the late time.
Thus, Eq.~\eqref{eq:restimate} is an attractor solution for the core radius.
}
\label{fig:CNFW}
\end{figure*}
We assume that semi-annihilation and self-interaction of DM play no role during the process of halo formation, and use the NFW density profile as the initial halo profile.
In the case of pure SIDM, using the initial NFW profile has been empirically checked to be in good agreement with the $N$-body simulations even in the case that the gravothermal collapse of a halo~\cite{Koda:2011yb,Essig:2018pzq} occurs.
On the other hand, this may not be the case in the presence of semi-annihilation of DM.
Since the $N$-body simulations for SHDM is not available at the moment, we cannot explicitly check if using the initial NFW profile would reflect the realistic evolution of a SHDM halo.
Instead, we check if the resultant SHDM halos are sensitive to the initial halo profiles in our fluid model.

In a cosmological simulation for SHDM, we expect that the heat injection from semi-annihilation would make the initial halo profile to be cored than the NFW profile.
Thus, we use the cored NFW (CNFW) profiles as different initial conditions; $\rho_{\rm CNFW}=\rho_s r_s/[(r_c+r_s)(1+(r/r_s))^2]$, where $r_c$ is the core radius, and $r_s$ and $\rho_s$ are the NFW scale parameters.
We obtain solutions to Eqs.~\eqref{eq:gravothermaleqns} for different values of $r_c$ for the initial CNFW profile.
The results are summarized in FIG.~\ref{fig:CNFW}.
The right panel of FIG.~\ref{fig:CNFW} is the time evolution of the core radius of a field dwarf/LSB galaxy for different values of the initial core radius $r_c$ (black lines).
We find that the black curves eventually converge to a single red line, which is the analytic fit for the core radius in the $r_c=0$ case:
\begin{equation}
\begin{aligned}
r_{\rm core}(t)\simeq 1.4\times r_s \left(\frac{t-t_\ast}{t_{\rm heat}}\right)^{0.57}\,.
\end{aligned}
\label{eq:restimate}
\end{equation}
Thus, we see that Eq.~\eqref{eq:restimate} is an attractor solution in time.
The left panel of FIG.~\ref{fig:CNFW} is a contour plot for the resultant core radius.
There, we see that the resultant core radius is insensitive to $r_c$ as long as $r_c\lesssim r_{\rm core}(t=t_{\rm age})$ (region above the red solid line).
We confirm that the benchmark parameter region for SHDM ($\left\langle\sigma_{\rm semi}v_{\rm rel}\right\rangle=6\times 10^{-26}\,{\rm cm^3/s}$, $\sigma_{\rm self}/m=0.1\,{\rm cm^2/g}$, and $m\sim 0.9\,{\rm MeV}$ with $b=3$) is unchanged as long as $r_c/r_s\lesssim0.1$.
\\
\paragraph{\underline{Comparison with $N$-body simulations}.---}
\begin{figure*}
\centering
\includegraphics[scale=0.55]{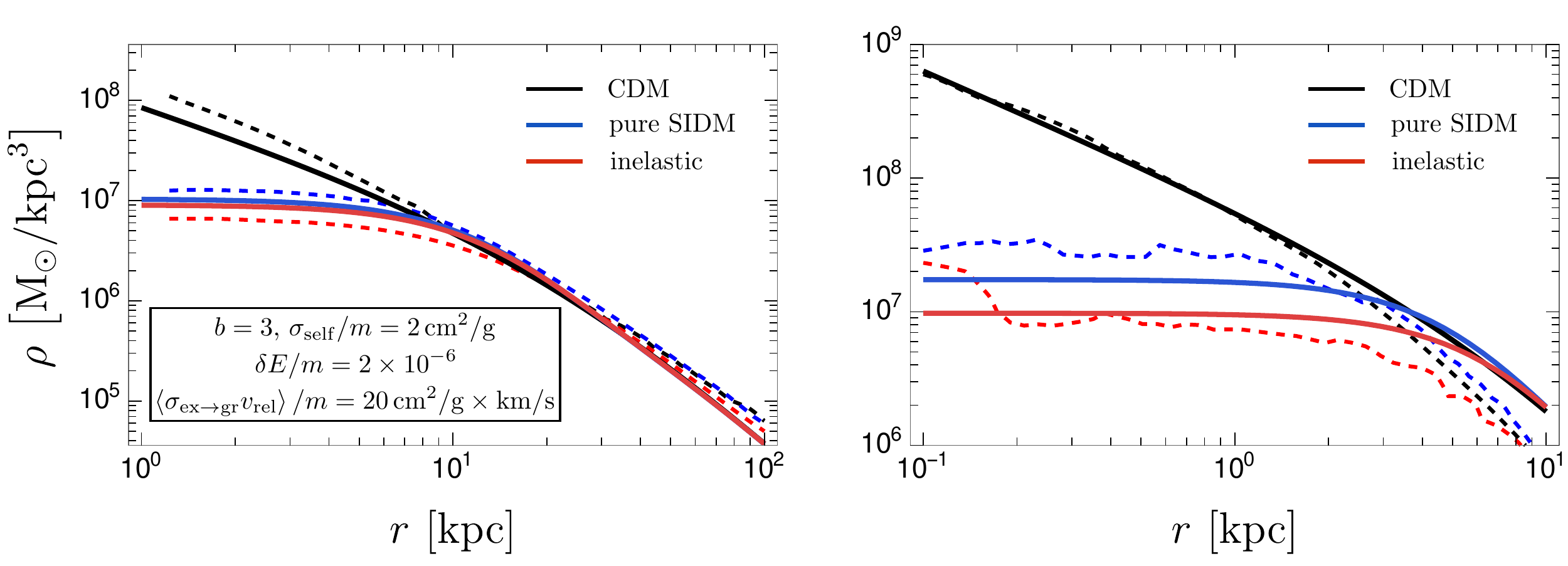}
\caption{
{\it Left panel} - Comparison between the resultant density profile of a MW-sized halo ($M_{200}=1.6\times10^{12}\,{\rm M}_\odot$) from the fluid model (solid) and the $N$-body simulation for the inelastic SIDM~\cite{Vogelsberger:2018bok} (dashed).
For the fluid model, the NFW profile ($\rho_s=0.0032\,{\rm M_\odot/pc^3}$, $r_s=31.1\,{\rm kpc}$) is used as the initial condition (solid black).
Red (blue) curves represent the case of inelastic (pure) SIDM.
{\it Right panel} - Same as the left panel but for a subhalo inside the MW-sized host halo. The dashed line corresponds to the median profile of the ten most massive subhalos from the $N$-body simulation.
For the fluid model, we use the NFW-fit ($\rho_s=0.0057\,{\rm M_\odot/pc^3}$ and $r_s=11.1\,{\rm kpc}$, which corresponds to $M_{200}=1.5\times10^{11}\,{\rm M_\odot}$) to the central CDM profile from the $N$-body simulation.
We set $b=3$ so that the core density from the fluid model matches the $N$-body simulation.
}
\label{fig:XDM}
\end{figure*}
We model the DM self-heating under the assumption that DM particles are heated by capturing the DM particles produced from exothermic DM scattering through DM self-interaction.
This introduces a fudge factor $b$ that represents the uncertainty in the capture efficiency $\xi$ (see Eq.~\eqref{eq:xi}).
We set the fudge factor to be $b=3$ as a reference value, so that our fluid model reproduces the core density from the $N$-body simulation for the inelastic SIDM~\cite{Vogelsberger:2018bok} (see the right panel of FIG.~\ref{fig:XDM}).

We remark that using the modeling for the heating [Eq.~\eqref{eq:semiinjection}] for the case of inelastic SIDM may be a naive application.
The difference from the case of SHDM is that the exothermic DM scatterings are frequent.
For example, inside MW satellites, the DM particles undergo at least one exothermic scattering during the age of the Universe, producing DM particles with considerable kick velocities, $\sqrt{\delta E/m}c\sim{\cal O}(100)\,{\rm km/s}$.
Since the capture efficiency is tiny [Eq.~\eqref{eq:xi}], most of the kicked DM particles will escape from the halo.
Such evaporation of DM particles may not be negligible, and we would need to incorporate an additional equation to Eqs.~\eqref{eq:gravothermaleqns} to take into account the evaporation of DM.

\end{document}